\documentclass[fleqn,usenatbib]{mnras}

\usepackage{newtxtext,newtxmath}
\usepackage[T1]{fontenc}

\DeclareRobustCommand{\VAN}[3]{#2}
\let\VANthebibliography\thebibliography
\def\thebibliography{\DeclareRobustCommand{\VAN}[3]{##3}\VANthebibliography}

\usepackage{graphicx}	% Including figure files
\usepackage{amsmath}	% Advanced maths commands

\usepackage{xspace}
\usepackage{xcolor}
\usepackage[normalem]{ulem}

\defcitealias{scowcroft16}{S16}

\title[CO and CN in Cepheids]{Near-IR CO and CN in classical Cepheids}

\author[S. Call et al.]{Scott G. Call,$^{1,2}$
Thomas Griffith,$^{4}$
Eric G. Hintz,$^{1}$
Steve Ardern,$^{3}$
Victoria Scowcroft,$^{3}$ 
Jared Davidson,$^{1}$\newauthor
and Benjamin Boizelle$^{1}$
\\
$^{1}$Department of Physics and Astronomy, Brigham Young University, Provo, Utah, 84602, USA\\
$^{2}$Department of Physics and Astronomy, University of Iowa, Iowa City, Iowa, 52317, USA\\
$^{3}$Department of Physics, University of Bath, Claverton Down, Bath, BA2 7AY, UK\\
$^{4}$Department of Astronomy, University of Maryland, College Park, Maryland, 20742, USA
}

\date{Accepted XXX. Received YYY; in original form ZZZ}

\pubyear{2025}

\begin{document}
\label{firstpage}
\pagerange{\pageref{firstpage}--\pageref{lastpage}}
\maketitle

% Abstract of the paper
\begin{abstract}
We present medium resolution near-infrared spectral measurements of the carbon monoxide (CO) and the cyano radical (CN) features in 12 Galactic classical Cepheids. The pulsation periods of our sample range from 5.5 to 69 days, and the stars studied each had five or more near-IR spectral observations. The CO and CN measurements were used to probe CNO abundances of these stars, and elemental abundance values from the literature were used to identify the trends of [C/N] and [O/N] with CN and CO. To put these measurements in context, we performed stellar atmosphere fitting to obtain estimates of stellar parameters, with a primary focus on effective temperature. Our measurements and temperature estimates show that CN is significantly affected by dredge-up of processed material. We provide discussion as to the potential nature of the recently confirmed classical Cepheid, ET~Vul, and connect our near-infrared CO measurements to the mid-infrared period-colour-metallicity relation. 
\end{abstract}

% Select between one and six entries from the list of approved keywords.
% Don't make up new ones.

%%%%%%%%%%%%%%%%%%%%%%%%%%%%%%%%%%%%%%%%%%%%%%%%%%

%%%%%%%%%%%%%%%%% BODY OF PAPER %%%%%%%%%%%%%%%%%%

\section{Introduction}
\label{sec:intro}

Classical Cepheid variable stars (hereafter, Cepheids) are supergiants in the process of crossing the instability strip (IS) and experience pulsations as a result. These stars are standard candles and play a crucial role in the cosmic distance ladder, as their pulsation period is directly related to their luminosity (i.e. the Leavitt Law) and can be used to determine distances and therefore the expansion rate of the universe. The Leavitt Law has been shown to have less dispersion at redder wavelengths \citep{madore91}, and as a result the near- and mid-infrared (near-IR and mid-IR, respectively) have become priority regimes for studying Cepheids. The advent of the \textit{James Webb Space Telescope} (\textit{JWST}) has enabled deeper studies of Cepheids at these wavelengths \citep[e.g.][]{riess24,freedman25}.

Infrared studies of Cepheids have had a dramatic impact on the distance ladder error budget by reducing the effects of extinction and intrinsic Leavitt law dispersion. One of the remaining contributors to the dispersion is expected to be metallicity. Decades of study \citep[e.g.][]{Freedman1990, Scowcroft2009,Romaniello2022} have been invested in understanding the metallicity effect; however, it remains an unresolved problem. \citet{riess22} estimate that Cepheid metallicity currently contributes 0.5\% to the LMC-anchored $H_0$ measurement. The most significant factor is the limited number of Cepheids for which direct metallicity measurements are available. Most previous metallicity studies on Cepheid populations beyond the Magellanic Clouds have relied upon indirect estimates such as metallicity gradients from H\textsc{ii} regions. \citet{scowcroft16} used mid-infrared photometry to show that the temperature dependent variations of carbon-monoxide (CO) have a significant effect on the mid-infrared colour of Cepheids, and suggested that this could be used as a metallicity indicator. In this work, we use time-series spectroscopic observations of a sample of Galactic Cepheids to study how CO varies through the pulsation cycle, and to investigate whether CO variations can be linked to metallicity.

Cepheids also provide important insights into stellar evolution via the study of their pulsations. For example, information about stellar interiors can be obtained through comparison of observational data with stellar evolution models, the Leavitt Law allows for calibration of mass-luminosity relationships, and long-term photometric programs can provide information about rate of period change that is an indicator of the direction of instability strip crossing. Time-series spectroscopic measurements are used to derive abundances and determine spatial and pulsational velocities as well as photospheric temperatures throughout phase. Time-series spectral measurements of Cepheids in optical wavelengths are common in the literature \citep[such as][]{veloce24}, however, time-series spectra in the near-infrared are relatively scarce.

The recent near-IR spectroscopic studies of Cepheids in the literature have been focused on stellar parameters (e.g. effective temperatures, surface gravities, metallicities) and not the changes in molecular species. \citet{inno2019} used medium resolution, single and double epoch $J$--band spectra to characterize five newly discovered, obscured Cepheids. Single epoch medium resolution spectra were used by \citet{minniti20} to confirm 30 classical Cepheids on the far side of the galactic disk. \citet{kovtyukh22} reported emission of He~I in the near-IR at 1.083-$\rm \mu m$ from X~Cygni at multiple phases. \citet{kovtyukh23} performed time-series high resolution spectroscopy of six classical Cepheids for the purpose of calibrating line depth ratios with temperature in the $H$--band.

In \citet{call24}, we presented time-series, medium-resolution near-IR spectroscopic measurements of CO absorption for one Cepheid, CP~Cephei. We build on that paper here with observations of 11 additional Cepheids, as well as four new observations of CP~Cep. In addition to CO measurements, we also measured the cyano radical (CN) feature at 1.1-$\mu \rm m$ and two hydrogen lines in the Paschen series, Paschen beta (Pa$\mathrm{\beta}$) and delta (Pa$\mathrm{\delta}$). CO and CN are temperature-dependent species that are cyclically
formed and destroyed in Cepheid atmospheres due to temperature
changes induced by the pulsation of the star. The motivation for this study is to provide a framework for interpretation of CO and CN in Galactic Cepheids, particularly because these molecular features are measurable at low resolutions (e.g. \emph{JWST}), whereas atomic features typically require higher resolving power.

A few stars in our sample have periods within or near the Hertzsprung progression (HP) of roughly 6-16 days, and the characteristic bump can be seen in the light curves of these Cepheids (see Appendix \ref{app:phase}). Where the bump is located on the light (or radial velocity) curve is dependent on the pulsation period: for stars with periods $\lesssim$10 days the bump will be on the descending portion of the light curve and on the ascending portion for longer periods. The main contributor to this progression in the classical Cepheids is a resonance between the fundamental and second overtone modes of pulsation \citep{simonschmidt76, buchler86, moskalik92}. These stars, like other pulsators, are prone to shocks which originate from the layers that drive pulsation. While not required for shock development, stars exhibiting the HP may experience amplified shocks.

The Cepheids in this work have periods between 5.5 and 69 days. For periods in the range $6~\text{d}\leq P \leq~60~\text{d}$, we expect to see a trend of increasing CO absorption with increasing period \citep{scowcroft16}. For the shortest period Cepheids in our sample, we expect very little CO absorption, as these stars spend the majority of their pulsation cycle at temperatures too high for CO formation. Our longest period target, S~Vul, lies in in the poorly understood ``long period inversion'' region of the CO-period relation seen in \citet{scowcroft16} but is included as the only long-period Cepheid visible in the northern hemisphere.  

We present time-series near-IR measurements and analyses of the temperature-dependent species, CO and CN. In Section \ref{sec:observations} we detail the instrumental set up and the parameters of our Cepheid sample. We describe the feature measurements and fitting routine in Section \ref{sec:analysis} and provide the results of both in Section \ref{sec:results}. A broader discussion of the results is given in Section \ref{sec:discussion}, and, finally, a summary is provided in Section \ref{sec:conclusions}.

\section{Observations}
\label{sec:observations}

The near-IR spectroscopic observations were obtained using the Astrophysical Research Consortium 3.5-m telescope at Apache Point Observatory (APO) in New Mexico, USA. Observations of Cepheids ran from July 2023 to April 2025. The near-IR spectrograph used, a \emph{TripleSpec} system, spans a range of $0.95 {\text -} 2.46\, \mathrm{\mu m}$ with a resolving power of $\mathrm{R} = 3500$ \citep{wilson04}. Typical near-IR spectra acquisition protocol was followed, including dithering the objects between two points on the slit, and observing A0V spectral type stars at similar times and airmass, $X = \mathrm{sec}(z)$ with $z$ being the zenith angle, for calibration and telluric correction. Most of the Cepheid observations have differences in airmass of $|\Delta X| < 0.1$. Spectral extraction, calibration, and telluric correction were performed using a modified version of \emph{SpexTool} \citep{cushing04, vacca03}.

Details of the Cepheids used in this work are presented in Table \ref{tab:obs}. The sample presented here is limited to Cepheids with five or more spectral epochs, and the total number of epochs is 115.  We adopt the periods from the \textit{Gaia} Data Release 3 \citep{gaia23}, and maximum light epochs were determined using photometry from the \textit{Transiting Exoplanet Survey Satellite} (\textit{TESS}) \citep{tess15} except in the cases of the longer period stars S~Vul, ET~Vul, and GY~Sge. For S~Vul, the \textit{Gaia} epoch was used. For ET~Vul and GY~Sge, we adopt the maximum light epoch from the better-sampled All-Sky Automated Survey for Supernovae (ASAS-SN) light curves \citep{shappee14,kochanek17}.
In the following sections, the phases of our spectral observations, $\phi$, were calculated using the epochs and periods in Table \ref{tab:obs} such that $\phi = 0$ corresponds to maximum light in the reference light curves. The three sources for the optical photometry cover different wavelength ranges; \emph{Gaia} $G$ covers most of the optical and peaks near 600-nm, ASAS-SN photometry is in $V$, and TESS covers $R$ to $I$. However, the phase offset between these bands is sufficiently small that it would not affect the analysis presented in this work.
 
Metallicity values listed in Table \ref{tab:obs} are from \citet{genovali14} with the exception of ET~Vul, where we adopted $\mathrm{[Fe/H]} = -0.08 \, \mathrm{dex}$ from \citet{trentin24}. It is worth noting that a previous estimate for ET~Vul was $[\mathrm{A/H}]=-0.4\,\rm dex$ from \citet{harris81} (the author uses heavy element abundance $[\mathrm{A/H}]$). The values listed in Table \ref{tab:obs} were used in performing LTE atmosphere fitting as described in Section~\ref{sec:analysis}.

\begin{table}
    \centering
    \begin{tabular}{c|c|c|c|c|c|c}
        Star & Period & Epoch HJD & $\langle J \rangle$ & $N$ & [Fe/H] &  Ref \\
             &   d    & +2400000 & mag & &dex & \\
         \hline
        RZ~Gem  & 5.52974  & 59517.370 & 7.607 & 5  & -0.16 & 1,3\\
        V359~Cam& 6.56972  & 59928.693 & 9.479 & 9  & –0.19 & 2,3 \\
        RS~Ori  & 7.56695  & 59543.680 & 6.411 & 7  & 0.11  & 1,3\\
        MN~Cam  & 8.17956  & 59929.242 & 8.140 & 11 & -0.02 & 2,3\\
        RY~Cas  & 12.13703 & 59874.030 & 7.063 & 12 & 0.32  & 1,3\\
        CP~Cep  & 17.86737 & 59829.030 & 7.330 & 13 & 0.05  & 1,3\\
        IY~Cyg  & 21.76763 & 60348.526 & 9.324 & 10  & -0.12 & 2,3\\
        BM~Per  & 22.96594 & 59914.050 & 6.656 & 14 & 0.20  & 1,3\\
        OT~Per  & 26.11064 & 59916.270 & 8.712 & 13 & -0.10 & 1,3\\
        GY~Sge  & 51.56623 & 57212.930 & 5.530 & 8  & 0.26  & 1,3\\
        ET~Vul  & 53.37459 & 57285.785 & 8.930 & 5  & -0.08  & 2,4\\
        S~Vul   & 69.46742 & 56824.005 & 5.410 & 8  & 0.09  & 1,3 
    \end{tabular}
    \caption{Cepheids in our sample with $N\geq 5$ spectral epochs. Periods are from \textit{Gaia} DR3; $\langle J \rangle$ references: 1 -\citet{Monson_2011}, 2 - \citet{skrutskie06}, [Fe/H] references: 3 - \citet{genovali14}, 4 - \citet{trentin24}.}
    \label{tab:obs}
\end{table}

\section{Analysis}
\label{sec:analysis}

\subsection{CO and CN measurements}

\begin{figure}
    \centering
    \includegraphics[width=\columnwidth]{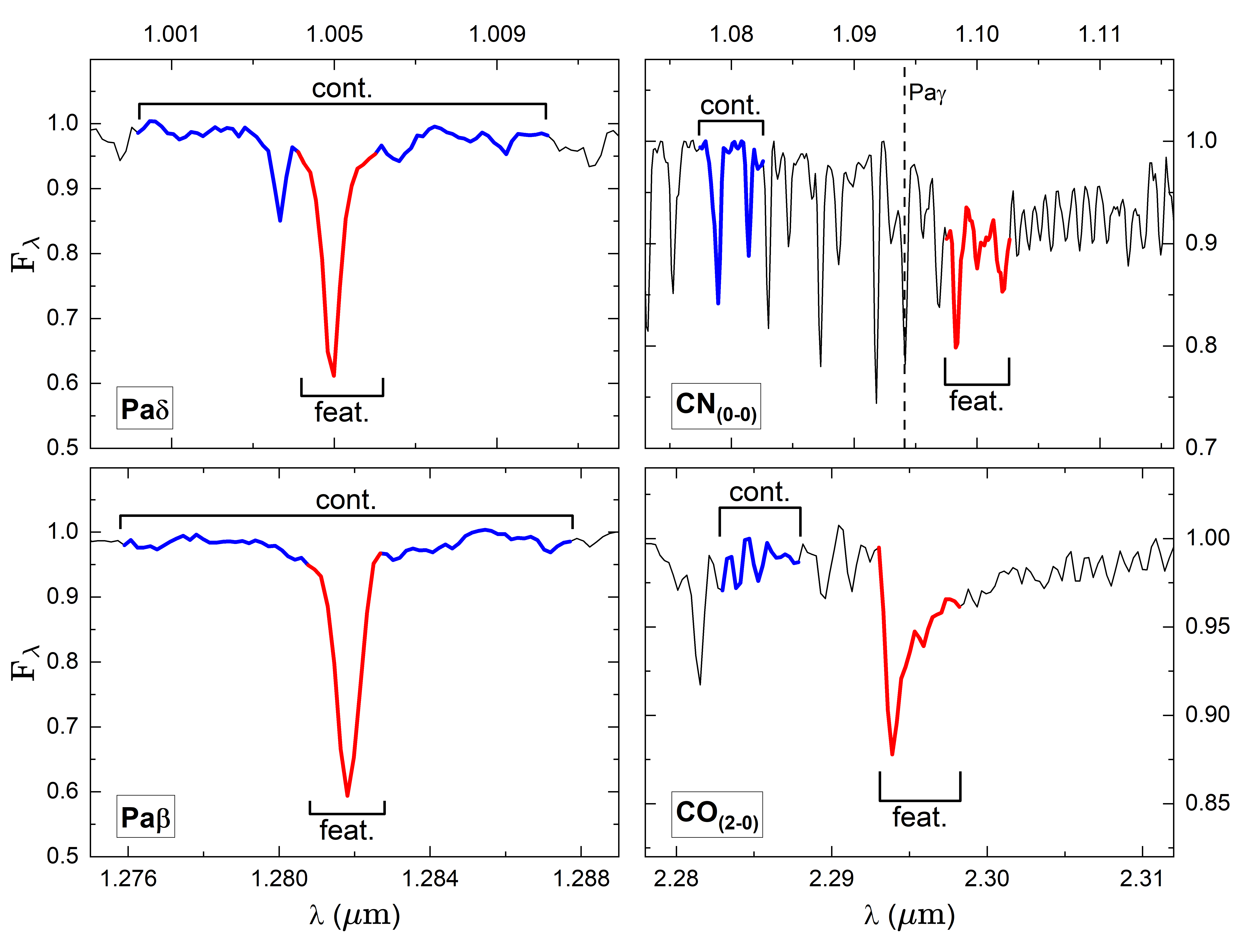}
    \caption{Examples of the indices in Table \ref{tab:bandmeasurements}, with the two Paschen line index regions are on the left and CN and CO regions on the right. These example spectra come from observations of BM~Per, with the left panels at phase $\phi = 0.01$ and the right panels at a cooler phase, $\phi = 0.88$. Note: these examples have been normalized for illustration purposes, while the index measurements were performed without normalization.}
    \label{fig:measurementrange}
\end{figure}

To quantify the abundances of CO and CN, we adopt the indices from \citet{kleinmann86} and \citet{gonneau16} for 2.3-$\mathrm{\mu m}$ CO and indices adapted from photometric filters from \citet{wing1971} for 1.1-$\mathrm{\mu m}$ CN. 
The adaptation of the index from \citet{wing1971} comes in the form of decreasing the width from 60~\r{A} to 52~\r{A}, for the sake of comparison with the CO index. In addition to CO and CN, we also measured two hydrogen features, the Pa$\mathrm{\beta}$ and Pa$\mathrm{\delta}$ lines, with both measurements made using on-line wide and narrow integrated flux similar to H$\alpha$ and H$\beta$ systems \citep[e.g.][]{straussducati1981}. These lines were chosen as they are indicative of temperature, with the strongest hydrogen absorption occurring at the highest temperatures. Pa$\mathrm{\beta}$ is relatively isolated near the edge of the $J$--band, and Pa$\mathrm{\delta}$ resides in the much busier 1.0--$\mathrm{\mu m}$ region. The other Paschen line in the spectra is Paschen Gamma (Pa$\mathrm{\gamma}$) at 1.094 $\mathrm{\mu m}$, but we did not include it as it is contaminated by the formation of CN at lower temperatures (though the hydrogen line does not affect our measurements of CN as the index regions are far enough removed, see Figure \ref{fig:measurementrange}). After shifting to zero-velocity, index measurements were performed using the \textsc{sbands} package of \textsc{iraf} \citep[][]{tody86, tody93}. The wavelength ranges for the features and corresponding continua are given in Table \ref{tab:bandmeasurements} and visual examples are shown in Figure \ref{fig:measurementrange}. Uncertainties were determined using the signal-to-noise (from root-mean square) of continuum regions near the features.

\begin{table}
    \centering
    \begin{tabular}{c|c|c|c|c}
        Line/Band & Feat. start & Feat. end & Cont. start & Cont. end \\
                  &$\mathrm{\mu m}$&$\mathrm{\mu m}$&$\mathrm{\mu m}$&$\mathrm{\mu m}$ \\
        \hline
        $\mathrm{CN_{(0-0)}}$& 1.0974&1.1026 & 1.0774&1.0826\\
        $\mathrm{CO_{(2-0)}}$& 2.2931&2.2983 & 2.2827&2.2879\\
        Pa$\beta$ & 1.2811 & 1.2831 & 1.2758 & 1.2878\\
        Pa$\delta$& 1.0042 & 1.0062 & 1.0002 & 1.0102  
    \end{tabular}
    \caption{Index measurement ranges. The CN ranges are adapted from photometric filters described in \citet{wing1971}, and the CO feature range is from \citet{harris81} while continuum is from \citet{gonneau16}.
    }
    \label{tab:bandmeasurements}
\end{table}

\begin{figure}
    \centering
    \includegraphics[width=\columnwidth]{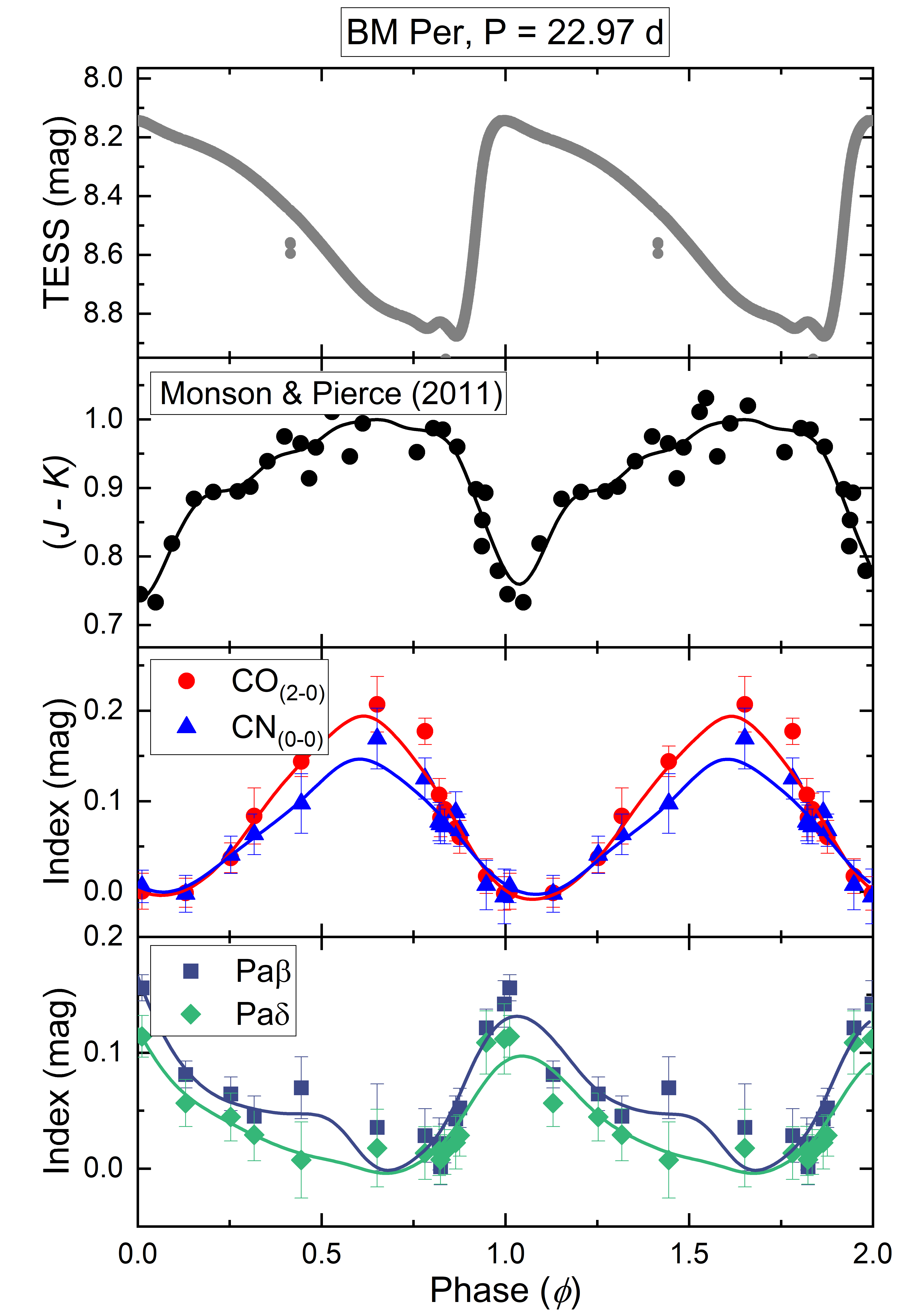}
    \caption{Measurements of CO, CN, and two Paschen lines compared with the TESS light curve and $(J-K)$ colour curve from \citet{Monson_2011} for classical Cepheid, BM~Per. The \textsc{gloess} \citep{persson04} curves are shown through the $(J-K)$ data and our index measurements.
    }
    \label{fig:phaseindex}
\end{figure}

Figure \ref{fig:phaseindex} puts observations of carbon species and Paschen lines in the Cepheid BM~Per in context with photometry in the near-IR and optical. The hydrogen lines reach maximum strength at the highest effective temperature, corresponding closely to peak brightness in the TESS light curve and the smallest difference in the $(J-K)$ color curve from \citet{Monson_2011}. Inversely, CO and CN are the strongest at cold temperatures as the optical light curve approaches minimum, or as the colour curve reaches a maximum. 

\subsection{LTE atmosphere fitting}
\label{sec:LTE}
To obtain effective temperature estimates for each observation, we used KORG, a spectral synthesis package for stars of spectral type F and cooler \citep{korg23, korg24}. KORG is able to interpolate MARCS model atmospheres for a range of stellar parameters and find the best-fit parameters for observational spectra. The user inputs initial estimates for effective temperature ($T_{\rm eff}$), surface gravity ($\mathrm{log}\, g$), microturbulent velocity, ($v_{\rm mic}$), and metallicity, prior to the fitting routine. In our fitting routine, we fixed the metallicity values to those from \citet{genovali14} for all objects besides ET~Vul, where we adopted the value from \citet{trentin24}, and the other three parameters were allowed to vary. We initially ran all observations through the routine with an initial parameter of $T_{\rm eff}=5500$~K. After this preliminary fitting, we then used the averages of the best-fit $T_{\rm eff}$ values for each star as the estimates for future fitting. 
For $\mathrm{log}\,g$, we used the period-gravity relation to determine the initial estimate \citep{tsvetkov1988}:
\begin{equation}
\label{eq:logg}
     \log g = -1.05\,\log P + 2.65 \, .
\end{equation}

It is worth noting that this is an estimate of \textit{mean} surface gravity, and that $\log g$ varies with pulsation phase. Scatter around the mean due to pulsation of $\Delta\log g \approx 0.8$ can be seen in work by \citet{andrievsky2002} and \citet{kovtyukh2005} show a similar range for their sample. The initial microturbulent velocity value was set to $v_{\rm mic} =3.5 \mathrm{\,km\, s^{-1}}$ for all cases. This estimate is on the lower end of expected $v_{\rm mic}$ \citep[see][]{kovtyukh2005}. Through testing of KORG with our data, we found that higher initial estimates (such as $v_{\rm mic} = 5\mathrm{\,km\, s^{-1}}$) would skew to unrealistically high final values more often. 

\begin{figure}
    \centering
    \includegraphics[width=\columnwidth]{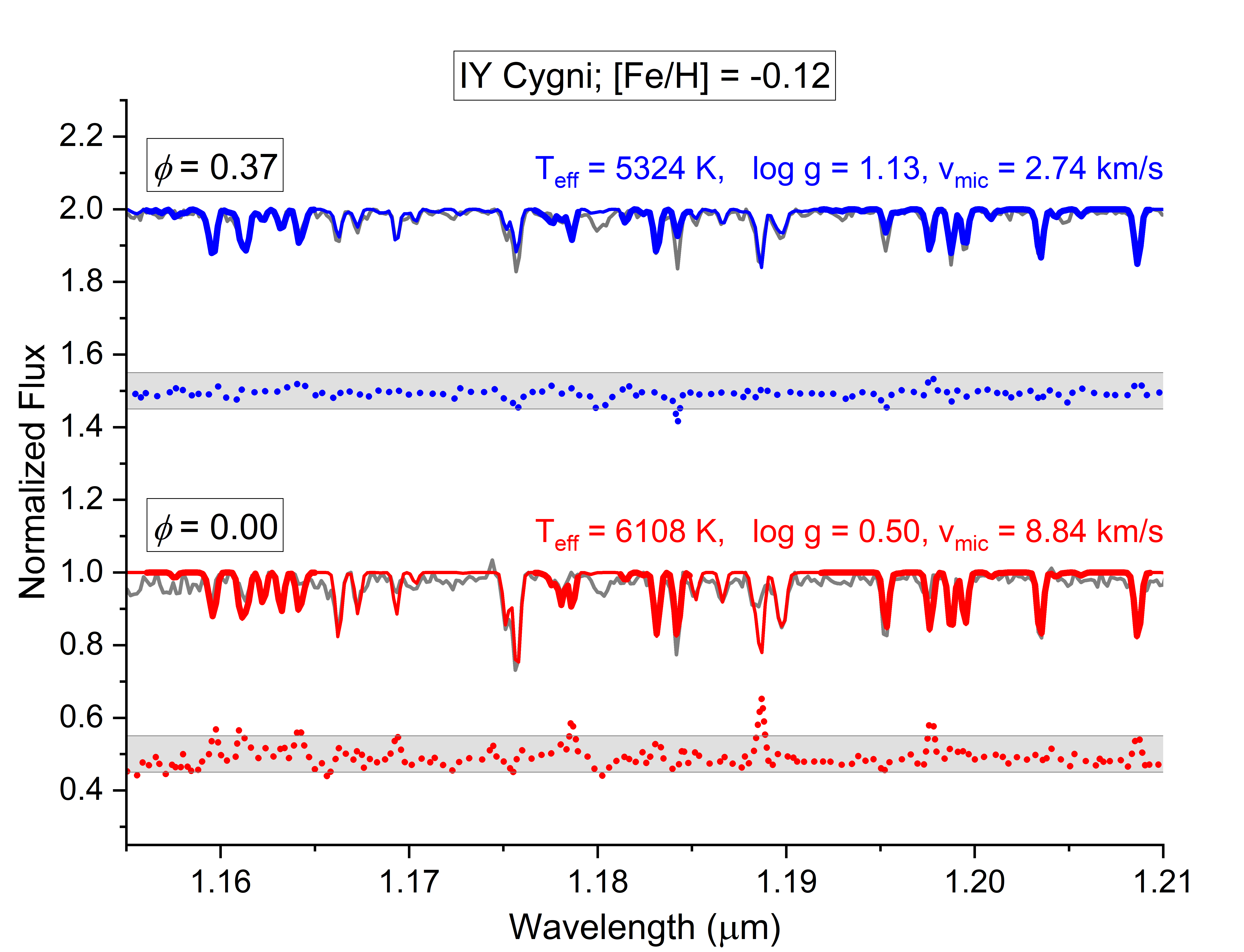}
    \caption{Spectral fitting via KORG for two observations of IY~Cyg. The gray lines are the observational data. The thicker lines overlaid are the fitting windows adapted from \citet{inno2019}, and the thin lines represent the synthesized spectra for the best-fit parameters. Residuals (offset from zero) are shown beneath both, where the shaded region represents $\pm 0.05$ or $\pm 5\%$.}
    \label{fig:fitting}
\end{figure}

Given that our spectra are medium-resolution, we chose to fit multiple atomic features in the $J$--band. We chose fitting windows similar to \citet{inno2019}, where species whose atmospheric abundance could be affected by mixing (such as C) were excluded. Prior to fitting, the observational spectra were normalized using the open-source python code, RASSINE \citep{cretignier20}. The goodness of the fit was based on reduced-$\chi^2$ values from the fit regions. Figure \ref{fig:fitting} shows spectra for two observations of IY~Cyg with the best-fit atmospheres overlaid and residuals. These spectra fits are representative of the sample, with one good and one poor fit based on the reduced-$\chi^2$ values. A potential cause of the poorly fit observations is changes in the atmospheric conditions, as the fitting region includes part of a strong telluric absorption region on the blue end. Small changes in the sky conditions between the target and the nearby standard star can also affect the correction of said absorption. On the other hand, a great fit does not necessarily indicate accurate parameters for the stellar atmosphere. In particular, the $\mathrm{log}\, g$ and $v_{\rm mic}$ occasionally extended beyond canonical values for Cepheid atmospheres. An example is shown in Figure \ref{fig:fitting}, where the best fit to the bottom (poor) observation having $\mathrm{log}\, g = 0.50$. The surface gravity is significantly lower than what one would expect given the star is close to the hottest point of its pulsation cycle. \citet{kovtyukh2005} determined stellar parameters for a Cepheid with a similar period, WZ~Sqr ($P=21.85$~d), and at its highest temperature ($T_{\rm eff} \approx 6200$ K) the surface gravity was $\mathrm{log}\,g \approx 1.8$. %

\section{Results}
\label{sec:results}
\subsection{Index Measurements}
\begin{figure*}
    \centering
    \includegraphics[width=\textwidth]{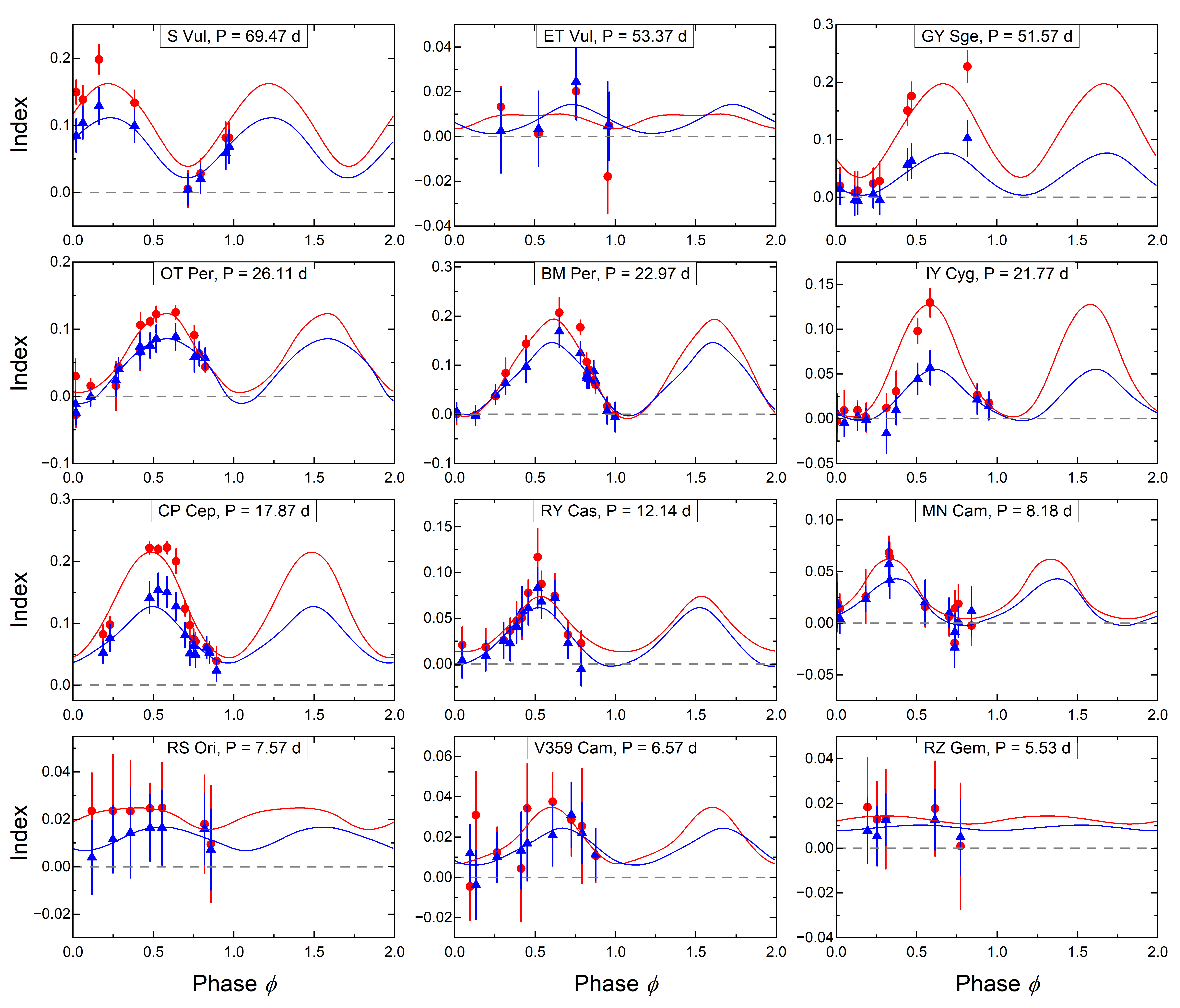}
    \caption{Phased CO and CN measurements for each star in the program, organized from top to bottom by period length (top-left longest, bottom-right shortest). Red circles represent CO and blue triangles, CN. \textsc{gloess} curves are shown following the same colour scheme, and continue through $1.0<\phi<2.0$ without our measurements for visualization purposes.}
    \label{fig:COandCN}
\end{figure*}

The CO and CN measurements for the stars in this study are shown in Figure \ref{fig:COandCN}, where the index magnitudes are plotted with phase of the star. Phase coverage of the stars varies with the longest period stars GY~Sge, ET~Vul, and S~Vul, and the shortest period star RZ~Gem, having the fewest observations. As expected, the shorter period stars exhibit weaker CN and CO features. RZ~Gem (P = 5.53 d) shows some indication of CO and CN absorption in the form of positive index values for both across phase, but zero is within standard uncertainties for the measurements. It is worth noting that the continuum regions are at shorter wavelengths than the feature regions, and a blackbody curve without any absorption features would yield slightly positive values for CO and CN indices as a result. V359~Cam and RS~Ori both have positive values for CO and CN, but they do not vary much throughout the pulsation phase. 

MN~Cam is the shortest period star (P = 8.18 d) in our sample that exhibits an obvious relation to pulsation phase indicated by significant CO and CN formation from $\phi =0$ to a maximum value at $\phi = 0.33$, followed by decrease in strength to near zero at later phases. 

RY~Cas and CP~Cep are sampled well through temperature minimum and reach higher index values than MN~Cam. The CN indices for CP~Cep are comparable to that of CO except at minimum temperature where the $(J-K)$ curve becomes flat \citep{call24}. The CO absorption is much stronger than CN at these phases, and both stop forming despite an extended period of colder $T_{\rm eff}$ (the three observations at maximum CO correspond to nearly two days). IY~Cyg does not appear to develop much CO or CN until $\phi \approx 0.4$, at which point there is a rapid increase in CO. The optical light curve of IY~Cyg features a slow decline from maximum to minimum light from $0.0 <\phi< 0.6$, at which point it remains at minimum light until $\phi = 0.9$. It then undergoes a very fast rise to maximum light in the last 10\% of its pulsation cycle (about 2 days), and we see the CO and CN abundances back near 0 at these phases. 

BM~Per and OT~Per are both well sampled throughout their pulsations. With a three-day longer period, we found OT~Per to have a similar mean effective temperature ($T_{\rm eff}=5352\pm100$~K) to BM~Per ($T_{\rm eff}=5429\pm100$~K).
While not as well sampled through the minimum temperature, BM~Per reaches stronger CO and CN measurements than OT~Per. This is due to their metallicities, as BM~Per is significantly more metal-rich than OT~Per.

Each of the stars with periods greater than 50 days have unique trends in their CO and CN measurements. GY~Sge reaches a similar CO maximum level to CP~Cep and BM~Per, while its CN value at max is about two thirds of that in CP~Cep. We have eight observations of the longest period Cepheid known in the northern hemisphere, S~Vul. Four of the observations are between $0.7 < \phi < 1.0$, and CN is comparable to CO in these. The other four are between phases 0 and 0.4, and in these the CO becomes significantly stronger than CN. ET~Vul is an interesting case, as the five observations span about 70\% of the phase but the CO and CN levels are near zero for all, and temperatures from KORG were above 6000 K for each observation. In Section \ref{sec:discussion}, we provide additional discussion of the abundances and potential nature of ET~Vul.

\begin{figure*}
    \centering
    \includegraphics[width=\linewidth]{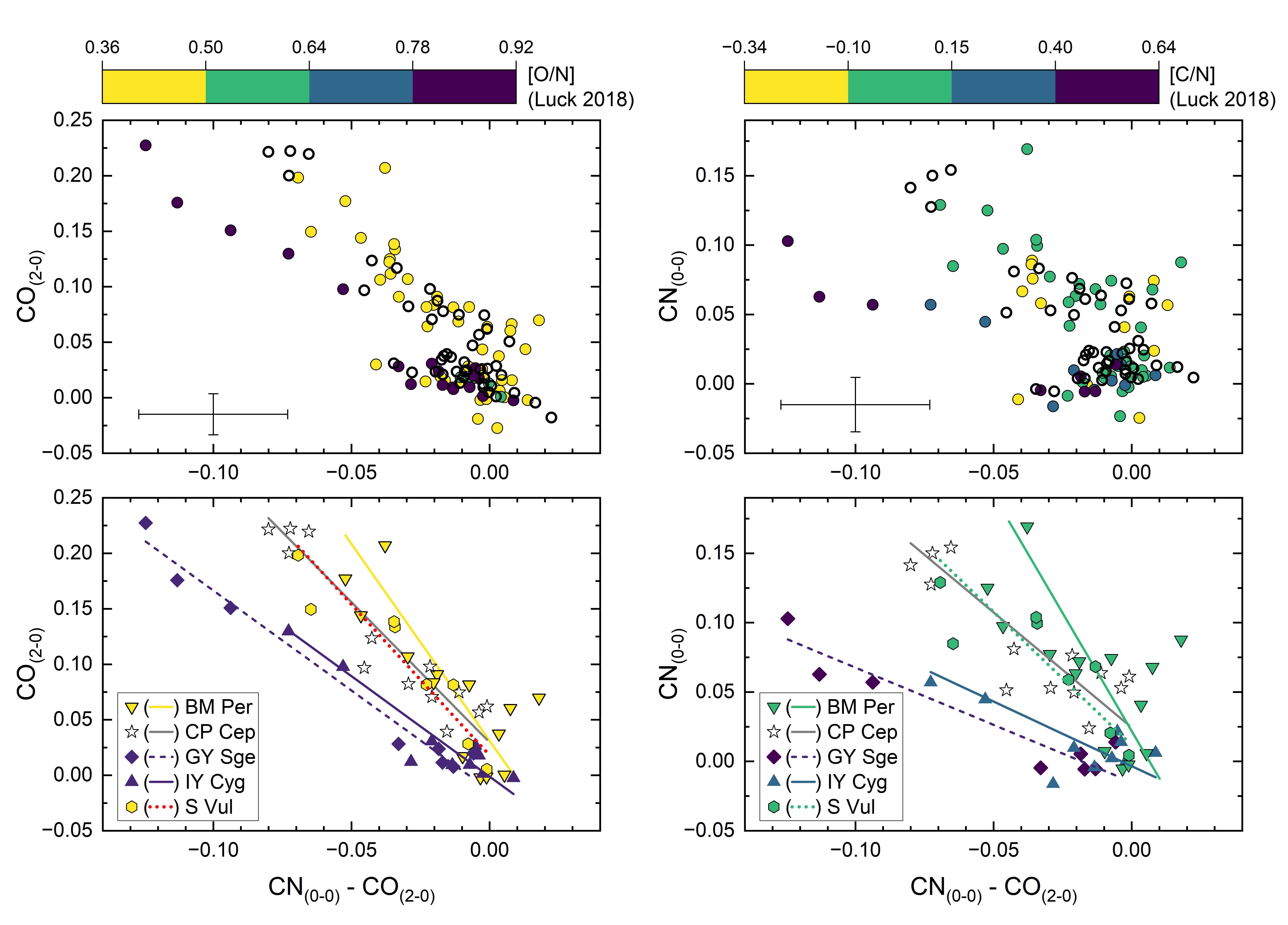}
    \caption{The $\mathrm{CO_{(2-0)}}$ (left panels) and $\mathrm{CN_{(0-0)}}$ (right panels) vs. $\mathrm{(CN_{(0-0)}}-\mathrm{CO_{(2-0)}})$. The filled observations colours represent stars with [C/N] (left) and [O/N] (right) derived by \citet{luck18}. Open circles are for stars without published N abundance. Mean errors are shown in the lower left corners of the top panels. The bottom panels are the same as the top but only for five stars:  BM~Per, CP~Cep, GY~Sge, IY~Cyg, and S~Vul. Lines represent weighted linear fits through the data for each star.}
    \label{fig:cno_luck}
\end{figure*}

In most observations, the stars exhibit higher CO absorption than CN, especially at the times of maximum strength of both, or at the lowest effective temperature. Notable stars with a large difference ($\sim0.06$) between CO and CN are S~Vul, GY~Sge, IY~Cyg, and CP~Cep. To probe these trends for our sample, we took the difference between CN and CO indices for each observation and plotted them against CO measurements in the top-left panel of Figure \ref{fig:cno_luck}, and against CN measurements in the top-right panel. In the former, we colourize these data based on average [O/N] abundances reported by \citet{luck18}. In the latter, a similar display is used for [C/N] abundances. Open circles correspond to stars for which \citet{luck18} did not have abundances of C or N. It is worth noting that the errors of our carbon species measurements are based on the SNR of estimated continua regions and are likely to be more on the conservative side, particularly for the cooler stars for which the continua are difficult to distinguish from metal lines. Both of the top panels feature a cluster at CO and CN $\approx0$ and another at slightly elevated values ($\approx0.5$ for both). All observations of shorter period stars (P < 8 d) in our sample are contained in the grouping near 0 as well as those for ET~Vul. Stars with periods greater than 8 days (excluding ET~Vul) will have observations in this cluster but also have measurements at higher CO and CN. 

The bottom panels of Figure \ref{fig:cno_luck} examine five stars: the four mentioned above with large differences between CN and CO strengths, S~Vul, GY~Sge, IY~Cyg, and CP~Cep, as well as BM~Per, which does not reach as large of a difference but reaches high values of both CO and CN. For each star, we performed a linear regression on the measurements at each phase which are shown as lines in the figures. As one would expect, the two stars with the highest [O/N] and [C/N] ratios, GY~Sge and IY~Cyg, form far more CO than CN. BM~Per and S~Vul have higher abundances of atmospheric N, leading to more CN formation and steeper slopes to the data. CP~Cep does not have published values for N, but comparing to the other stars' CO, CN, and abundance ratios we can provide estimates of its [O/N] and [C/N]. The similar slopes between data for CP~Cep and S~Vul suggest similar ratios for each star.

To go further, we found a relation between slope (CO or CN over difference) and the abundance ratios derived by \citet{luck18} for these four stars, and applied that relation to the slopes found for CP~Cep. We obtained [O/N]$\approx$0.60 and [C/N]$\approx$0.25 for CP~Cep. The same relation applied to the four stars with known abundances yielded [C/N] and [O/N] ratios within $\sim$$0.22$ and $\sim$$0.15$, respectively. While no replacement for high-resolution spectroscopic measurements, simultaneous measurements of these carbon features can provide an estimate of the CNO abundances relative to each other. 

\subsection{KORG Fitting}
\begin{table}
    \centering
    \begin{tabular}{c|cc|cc|cc}
        Star & $\langle T_{\rm eff}\rangle$ & S.E. & $\langle\log\,g\rangle$ & S.E. & $\langle v_{\rm mic} \rangle$ & S.E.  \\
        & (K)      &  & [cm s$^{-2}$] & & (km s$^{-1}$)&  \\
        \hline
        RZ~Gem  & 6071 & 45 & 2.31 & 0.22 & 5.3 & 0.4\\
        V359~Cam& 6055 & 33 & 2.24 & 0.17 & 6.9 & 0.3\\
        RS~Ori  & 6225 & 38 & 2.72 & 0.19 & 4.8 & 0.4\\
        MN~Cam  & 5891 & 30 & 1.86 & 0.15 & 5.7 & 0.3\\
        RY~Cas  & 5587 & 29 & 1.41 & 0.14 & 3.7 & 0.3\\
        CP~Cep  & 5170 & 28 & 1.57 & 0.14 & 4.0 & 0.3 \\
        IY~Cyg  & 5818 & 32 & 0.92 & 0.16 & 5.9 & 0.3\\
        BM~Per  & 5429 & 27 & 0.90 & 0.13 & 4.4 & 0.3\\
        OT~Per  & 5352 & 28 & 1.17 & 0.14 & 4.5 & 0.3\\
        GY~Sge  & 5331 & 35 & 0.60 & 0.18 & 5.0 & 0.4\\
        ET~Vul  & 6331 & 45 & 1.20 & 0.22 & 11.6& 0.4\\
        S~Vul   & 5798 & 35 & 1.29 & 0.18 & 8.5 & 0.4 
    \end{tabular}
    \caption{Mean parameters from KORG model fitting. S.E. is the standard error of the mean ($\sigma/\sqrt{N}$) for the preceding column. The following were adopted for the individual measurements:$ \sigma_{T}=100$ K, $\sigma_{\log g}=0.5$, $\sigma_{v} = 1.0\, \mathrm{km\, s^{-1}}$.}
    \label{tab:korg}
\end{table}

The mean parameters from the KORG fitting routine for each star are given in Table \ref{tab:korg}. We did not exclude any observations (for poor fits, etc.) or implement any weights when calculating the mean. As a result, stars with better sampling are likely to have more accurate mean temperatures. For example, the OT~Per observations are well-spaced throughout its pulsation phase while 6 out of 10 observations of IY~Cyg are within the hottest third of the pulsation cycle, which will skew the mean temperature. 

\begin{figure*}
    \centering
    \includegraphics[width=\textwidth]{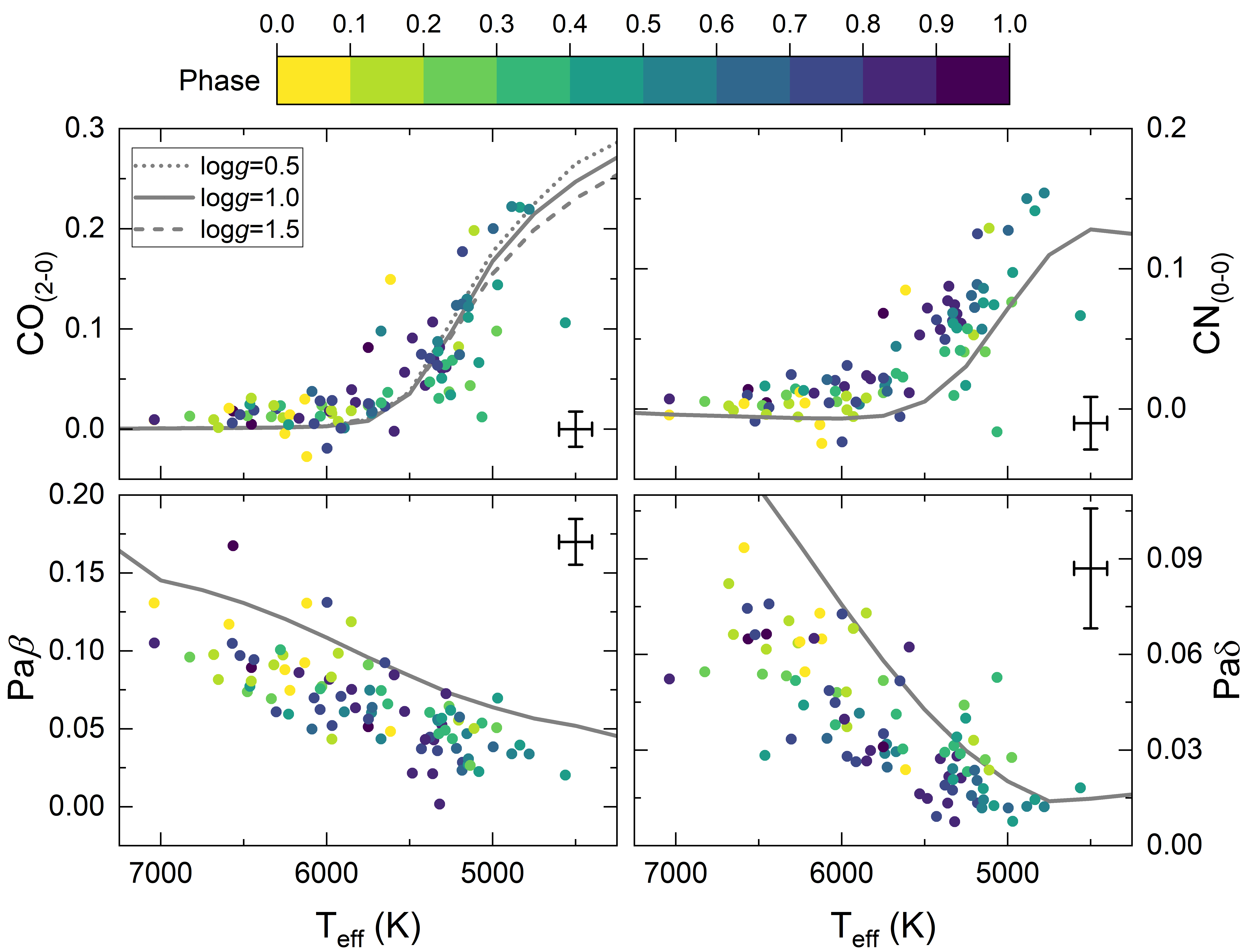}
    \caption{Measurements of CO, CN, Pa$\mathrm{\beta}$, and Pa$\mathrm{\delta}$ indices as well as effective temperatures from spectral fitting with KORG. The lowest quality fits were removed (23 out of 115 observations). Observations are coloured with respect to binned phase. Measurements of the same lines/bands were performed on synthesized spectra at solar metallicity, and are shown as lines in each panel. Mean errors are in corner of each panel.}
    \label{fig:temperaturecomparison}
\end{figure*}

The best-fit temperatures for our observations are shown in Figure \ref{fig:temperaturecomparison}, plotted with each of the index measurements. As mentioned in Section \ref{sec:analysis}, the goodness of fit was determined via reduced-$\chi^2$. For this analysis, we set a reduced-$\chi^2$ threshold such that we kept the best 80\% of observations which are shown in the figure. As expected, the hydrogen lines trend with temperature. The Pa$\mathrm{\beta}$ relation appears tighter due to its relative isolation compared to other lines and its greater strength in comparison to the Pa$\mathrm{\delta}$ line. The CO and CN relations are similar to each other. At higher temperatures, $T_{\rm eff} >  6000\, \mathrm{K}$, both CO and CN indices are effectively zero. At temperatures $T_{\rm eff} \lesssim 5500\, \mathrm{K}$, both carbon species form relationships with temperature. Similar work for stellar populations has shown the relationship between the CO band head and temperature, such as \citet{marmol-queralto08}. In their work, they determined higher-order relations and included metallicity as a parameter. Because our sample is centred on roughly solar values of metallicity and our fitting routine was not as robust, we elected to compare our measurements to synthetic MARCS atmospheres instead of empirical relations \citep{marcs08}. These atmosphere models were generated using KORG in 250 K increments at solar metallicity ([Fe/H]~=~0.0), and the same indices were measured using \textsc{sbands} as described in Section \ref{sec:analysis}. In the CO portion of Figure \ref{fig:temperaturecomparison}, the synthetic atmosphere measurements are shown for surface gravities $\mathrm{log}\, g =0.5,1.0,\mathrm{and\,}1.5$. In the other three panels, only $\mathrm{log}\, g = 1.0$ is shown, as the measurements varied little with surface gravity for our temperature range. For example, the CN measurement at 5000 K for the $\mathrm{log}\, g=1.5$ synthetic spectrum was 0.006 lower ($\sim8\%$) than the $\mathrm{log}\, g = 1.0$ model.

Indices for the two Paschen lines trend with temperature and follow the slopes of the synthetic models. However, the majority of our measurements are weaker than the model predictions. This is potentially due to the non-local thermodynamic equilibrium (NLTE) effects on the hydrogen lines. In many cases, the shapes of the Paschen lines in our spectra are asymmetric and weaker than the LTE models, and these effects vary with pulsation. Shocks cause at least some of the asymmetry seen in the hydrogen lines. Two examples of asymmetry in the Pa$\beta$ line are provided in Appendix \ref{app:asymmetry}. Because of this, we chose to avoid hydrogen lines (both Paschen and Brackett) in our fitting routine and to adapt similar fitting windows to \citet{inno2019}.

\begin{figure}
    \centering
    \includegraphics[width=\columnwidth]{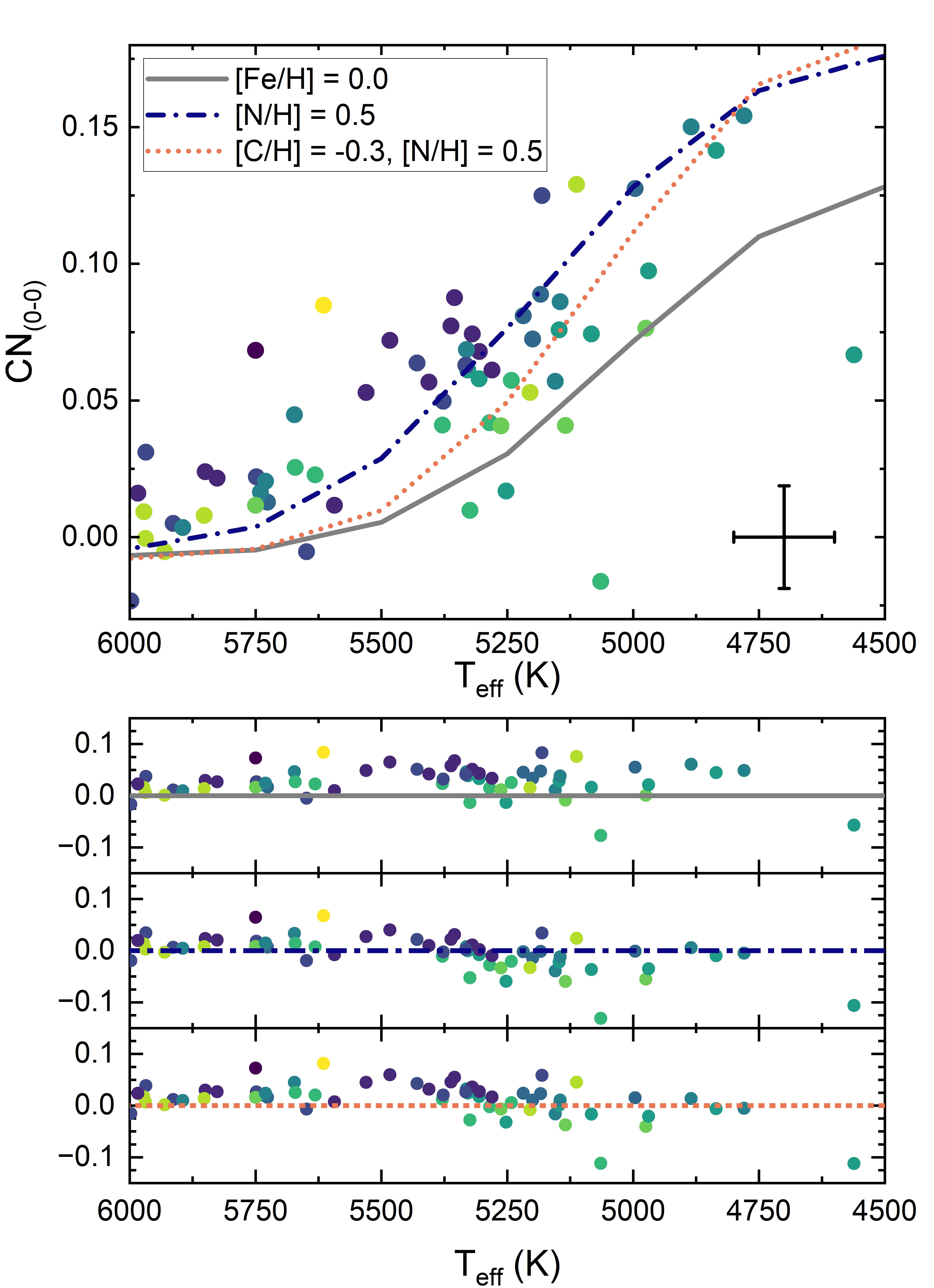}
    \caption{Zoom-in to the lower temperatures on the CN panel from Figure \ref{fig:temperaturecomparison}. Lines represent the measurements made on synthetic spectra for solar (gray solid), nitrogen-enriched (blue dashed-dot), and nitrogen-enriched with decreased carbon (orange dot) as described in the text. The residuals for each synthetic track are shown in the bottom panel in the same order. Colours of observations represent phase as in Figure \ref{fig:temperaturecomparison}.}
    \label{fig:CNtemp}
\end{figure}

The CO band head measurements follow the results from synthetic atmospheres very well, with the solar metallicity models nearly bisecting the observations between 5000 and 5500 K. CN, on the other hand, is significantly stronger in our observations than the standard, solar-metallicity atmospheric models. Nitrogen over-abundance in evolved supergiants is expected to be the result of the first dredge-up and/or rotational mixing during the main-sequence stage, with the former in better agreement with evolutionary timescales \citep[see][]{lyubimkov11}. In Figure \ref{fig:CNtemp}, we show the CN data zoomed in to the cooler temperatures. In addition to the solar-metallicity, $\mathrm{log\,}g = 1.0$ line, we also included two other model atmosphere tracks generated at $\mathrm{log\,}g = 1.0$. The first was synthesized using KORG with enriched N (to account for mixing), to [N/H] = 0.5 dex and all other elements left at solar values. This provided a visually better fit to our observational CN data than the solar abundance only model as shown in the residuals in Fig.~\ref{fig:CNtemp}. For the other additional model we kept [N/H] = 0.5 dex but also adjusted carbon such that [C/H] = $-0.3$ dex to compare with empirical results from \citet{kovtyukh96} and \citet{takeda13}. As seen in Fig.~\ref{fig:CNtemp}, these enriched N and decreased C models yield more CN than the strictly solar abundances case, particularly at lower temperatures. 

The colour bar in Figure \ref{fig:temperaturecomparison} represents the binned phases of the observations. The stars in the sample have different mean temperatures, but trends in phase can still be seen in the data. Phases near 0 or 1 are typically the hottest, and this can be seen with the yellow and light green colours mostly found between $6000\,\mathrm{K}< T_{\rm eff} < 7000\, \mathrm{K}$. Later phases are also expected to be approaching hotter temperatures, but it is typical for a star to have a slow decrease in temperature from $\phi > 0$ and a quick increase close to $\phi = 1$, hence the darker colours being more spread across $T_{\rm eff}$. Most stars reach maximum CO and CN at phases between $\phi \approx 0.4$ and $ \phi \approx 0.7$ (via Figure \ref{fig:COandCN}), and we would expect these phases to be when effective temperature is near minimum. The data shown in Figure \ref{fig:temperaturecomparison} are dominated by these phases at $T_{\rm eff} < 5500\, \mathrm{K}$. This is particularly interesting in the CO and CN panels of Figure \ref{fig:temperaturecomparison}. For CO, the observations below the synthetic measurement lines are generally at earlier phases than those above, and this can be better seen for CN in Figure \ref{fig:CNtemp} using the [N/H]=0.5 line as a reference. This may be indicative of a lag between the formation/destruction of CO and CN and temperature changes during pulsation. As the atmosphere expands and the $T_{\rm eff}$ reaches values where CO and CN can exist, the species slowly begin to form. When the atmosphere contracts and heats up, the CO and CN dissociate at a faster rate than they formed.

Appendix \ref{app:phase} includes the KORG fitting phase results for each star ($T_{\rm eff}$, $\log g$, $v_{\rm mic}$), the CO and CN measurements, optical light curves, and mid-IR $[3.6]-[4.5]$ colour curves for the two stars in our sample with \emph{Spitzer} photometric data \citep{monson12}.

\section{Discussion}
\label{sec:discussion}

\subsection{Effects of shocks on index measurements}

As mentioned in Section \ref{sec:results}, shocks contribute to the asymmetry seen in the Paschen lines. Shocks in Cepheids have been studied primarily via the hydrogen Balmer lines \citep[e.g.][]{gillet14, hocde20}. Unlike the n=2 lower level for the Balmer series, the n=3 levels of the Paschen series are not metastable and line formation will be closer to LTE. When shocks pass through the atmosphere, however, the collision rate increases and electrons in the overpopulated n=2 level can be excited to higher levels. These collisionally excited electrons can enhance the asymmetry effects of the shock observed in the Paschen lines. \citet{wallerstein19} found minimal effects of shocks on the Pa$\delta$ and higher lines. Likewise, we see small effects on the Pa$\delta$ and Pa$\gamma$ lines, but a much more significant effect on Pa$\beta$.

We have identified strong emission in the Pa$\beta$ line for two Cepheids in this sample, BM~Per and OT~Per (an example of the latter is shown in Figure \ref{fig:otper_shock}). Two observations of MN~Cam appear to show emission in Pa$\beta$ but not to the same extent as the other two. Less extreme profiles, such as steep absorption in the red or blue wings, appear in others (CP~Cep, IY~Cyg, RS~Ori, RZ~Gem, S~Vul, V359~Cam - Figure \ref{fig:asymm}). As one would expect, these asymmetric profiles yield weaker line measurements than the LTE case. The higher order Paschen lines also exhibit asymmetry but to lesser extent, and we would expect Pa$\alpha$ at 1.875-$\rm \mu m$ to have stronger effects than Pa$\beta$\footnote{Due to telluric absorption Pa$\alpha$ cannot be measured from ground-based observatories like APO.}. It is worth noting that Brackett-$\gamma$ (2.166-$\mathrm{\mu m}$) also experiences drastic changes in line shape.

As additional evidence of shocks in these stars, we also identified the He~I triplet emission at 1.083-$\rm \mu m$ first noted in classical Cepheids by \citet{kovtyukh22} and \citet{andrievsky23}. The authors identified phase-dependent emission of this helium feature in high resolution near-IR spectra for X~Cygni ($P=16.4$~d). They showed that the line velocity of this emission is consistent with a shock front moving through the atmosphere. Two Cepheids in this sample exhibit strong He~I emission at 1.083-$\rm \mu m$ at multiple phases: MN~Cam ($P=8.18$~d) and CP~Cep ($P=17.87$~d). Some observations are accompanied by increased absorption on the blue-side consistent with the P-Cygni profile. CP~Cep experiences very strong, blue-wing absorption, though it is worth noting that this absorption becomes blended with a relatively strong metal line (Si~I). MN~Cam is within the HP while CP~Cep is outside of the typical range, though it has a bump on the ascending part of the light curve. An interesting result is that the star with the strongest He~I emission, CP~Cep, does not exhibit hydrogen profiles nearly as striking, while the star with the most striking Pa$\beta$ profile, OT~Per, does not exhibit detectable He~I emission. We did not detect He~I emission in the other stars nor emission in the hydrogen lines, though each display some asymmetry in Pa$\beta$.

The profiles of Pa$\beta$ in our sample can be compared to more in-depth studies of shocks in the optical wavelengths. While comparing  H$\alpha$ profiles in X~Cyg to those in \citet{nardetto08}, \citet{gillet14} suggested that the primary shock reaches full development at different points in the atmosphere depending on the period. For classical Cepheids with $P\lesssim 17$~d, H$\alpha$ was not observed in emission. This was attributed to the shock reaching full development in the more dense (deeper) atmospheric regions, whereas H$\alpha$ emission was seen in some of the longer period stars. Likewise, we do not see Pa$\beta$ in emission in the stars with shorter periods with the exception of MN~Cam, but we do see emission in longer period OT~Per and BM~Per. More observations (ideally at higher resolution) of MN~Cam would build confidence that the emission is real and tied to propagation of shocks.

Any effects of shocks on the CO and CN features are less apparent than the hydrogen lines. We do not detect significant changes to absorption band shape for any observation. Since both species are dependent on temperature, we might expect to see increased dissociation due to the increased temperature of the shock wake – particularly in CN where the dissociation energy is lower than CO. It may also be true that because the CO and CN form during cooler times when the atmosphere is expanded much of the shock energy may have gone toward deeper H and He excitation prior to reaching the molecular population.

\subsection{CN, CO, and mixing}

The atmospheric abundances of C and N in Cepheids and other evolved, non-pulsating supergiants are different than the abundances in their main-sequence progenitors as a result of mixing \citep[see][]{lucklambert81,lucklambert85,kovtyukh96, takeda13}. Mixing is likely due to convection during the first dredge-up where the C-to-N-cycled material is pulled up to the higher layers of the star resulting in an overabundance of N and an underabundance of C. On the other hand, the abundance of O remains relatively unaffected by this stage of mixing.

The analyses of the species, CO and CN, provide an interesting perspective of the results of mixing. The most drastic elemental abundance change, N, is apparent when looking at CN vs. $T_{\rm eff}$ in Figures \ref{fig:temperaturecomparison} and \ref{fig:CNtemp}. The solar-abundance synthetic spectra severely underestimate the amount of CN formed in the atmospheres of these stars, while models generated with enriched N yield more reasonable estimates. Because O is essentially unchanged, the solar-abundance model spectra are adequate at estimating the CO strength at different temperatures despite C being underabundant.

Measurements of CO and CN in the near-IR can be used to estimate abundance ratios of the elements involved as discussed in the previous section (Figure \ref{fig:cno_luck}). This method requires the absorption of the carbon species to be sufficiently high (i.e. at temperatures of $T_{\rm eff}<5500$~K long enough for association), and the relation must be calibrated using stars with known CNO abundances. Measurements of the CO and CN features can be made with lower-resolution instruments. This method can also be applied to non-pulsating stars. In that case, only one observation (of the CO and CN regions) of the star would provide the information necessary to derive the abundance ratios. 

\subsection{ET~Vul - a candidate merger-channel Cepheid}
\label{sec:et_vul}
Our analysis shows ET~Vul is an outlier in terms of CO and CN in the photosphere, showing very low levels compared to the other Cepheids in our sample, even when compared to the other longer-period Cepheids S~Vul and GY~Sge. We also find ET~Vul to be an outlier in terms of both surface gravity and $T_{\rm eff}$. Indeed, ET~Vul's status as a classical Cepheid has only recently been confirmed by \citet{berdnikov20} through analysis of O-C Diagrams from 130 years of photometric data.  

We propose that these outlier properties could be indicative of a Cepheid formed through the binary merger mechanism described in \citet{2024A&A...686A.263P}. This would also result in a higher age than expected for a classical Cepheid. There is growing evidence that Cepheids may have a second non-canonical formation channel linked to the high-prevalence of finding them in binary and higher stellar systems. In observing Binary Double (BIND) Cepheids (two-Cepheid systems), \cite{2024A&A...686A.263P} noted that the ratios of pulsation periods of these BIND Cepheids do not conform to statistical expectations for pairs of coeval Cepheids. Together with inferred mass-differences between BIND Cepheids, they propose this is suggestive of a merger origin for Cepheids in more than 50\% of such systems.  Depending on the physical separation of system components, such a merger could take place during the MS, or post-MS for either one or both components.  \cite{2024A&A...690A.385D} performed simulations of the evolution of stars in the ZAMS mass range for Cepheids when found in multiple systems.  These simulations also suggest a second evolution channel for Cepheids. 
The statistical analysis of BIND Cepheids by \citet{2024A&A...686A.263P} and simulation work by \citet{2024A&A...690A.385D} taken together imply that a significant proportion of Cepheids may be much older than they appear, with a recent candidate merger-channel Cepheid having been identified with an age of ca. 1.1 Gyr \citep{Espinoza-Arancibia}.

The principal piece of evidence for a binary merger origin for ET~Vul would be a lower-than-expected surface gravity. Our mean surface gravity for ET Vul of $\langle \log g\rangle = 1.20$ is higher than the estimate from Equation \ref{eq:logg} of 0.8, but four out of five of the observations had KORG best-fit values $\log g < 1.2$. The KORG routine also returned $v_{mic}$ values much higher than most of the other stars in the sample. \citet{trentin24} report lower $\log g$ values for ET Vul, their highest value being $\log g$ = 0.58 ± 0.15 and this from higher resolution echelle spectra. Low $\log g$ could be indicative of a high rotation rate, itself driven by angular momentum transfer during mass transfer, a process understood to be very efficient \citep{2024ApJ...971...80S}. In this case we would expect higher centrifugal forces, a larger equatorial radius, and a longer pulsation period than might be assumed for its mass \citep{2014A&A...564A.100A,2016A&A...591A...8A}. However, \citet{trentin24} also report a projected rotational velocity (broadening velocity) that is typical for classical Cepheids which is hard to reconcile with the low $\log g$ values.

Even with a high $T_{\rm eff}$, we would anticipate an extended photosphere for ET~Vul due to its low surface gravity.  Hence the regions of the photosphere further from its base where we infer $T_{\rm eff}$, could be cool enough for the formation of CO and CN. Our observed low values of CO and CN could therefore be consistent with a complex co-evolutionary history for a merger-channel Cepheid, with the outer layers of the star being gained by mass transfer from a companion.  Potential scenarios where the transferred mass could be depleted in CNO elements include: hydrogen transferred prior to CNO dredge-up in donor, helium transferred from stripped red-giant core, helium transferred from post common envelope remnant, or fractionation of transferred mass in accretion disc.

ET Vul's $T_{\rm eff}$ is beyond the blue edge of the instability strip for a typical Cepheid. Modelling by \cite{2016A&A...591A...8A} showed the blue edge not to be significantly dependent on rotation rate, however they showed a weak dependence of the blue edge on metallicity, with low metallicity pushing the blue edge to higher $T_{\rm eff}$. Our finding of no CO or CN is suggestive of low metallicity, which is consistent with a high $T_{\rm eff}$. This lends further support to the idea that ET Vul has been the recipient of low-metallicity material from a donor star.

\subsection{Effect of CO on the period-colour relation}
\label{sec:co-pcz}

One of the main drivers for moving to the infrared when using Cepheids as distance indicators is the smaller intrinsic width of the Leavitt law (LL, previously known as the period-luminosity relation) at longer wavelengths. The LL can be considered a projection of the three-dimensional period--luminosity--colour relation (PLC) onto the two-dimensional period--luminosity plane \citep{madore91}. As such, moving to longer wavelengths where changes in temperature (hence colour) have minimal impact on luminosities should result in reduced dispersion ($\sigma$) in the LL. \citet{scowcroft11} and \citet{monson12} confirmed the reduced dispersion in the $3.6~\mu$m LL finding $\sigma \sim 0.10$~mag in [3.6], compared to $\sigma \sim 0.30$~mag in $V$. 
However, \citet{scowcroft11} found a small but significant increase in $\sigma$ at $4.5~\mu$m, a strong $[3.6]-[4.5]$ period-colour relation, and significant, non-zero $[3.6]-[4.5]$ colour variations throughout the pulsation cycle. \citet[][hereafter \citetalias{scowcroft16}]{scowcroft16} showed that the the cyclical colour variation and the period-mean-colour relation are both driven by the temperature-dependent $4.6~\mu$m CO feature. \citetalias{scowcroft16} found evidence that the mid-IR colour was dependent on both period and metallicity, suggesting that mid-IR colour has the potential to be used as a photometric metallicity indicator. In this section, we use our time-resolved observations to investigate this claim.

\begin{figure*}
    \centering
    \includegraphics[width=\textwidth]%{CO_logP_Teff_labelled.png}
    {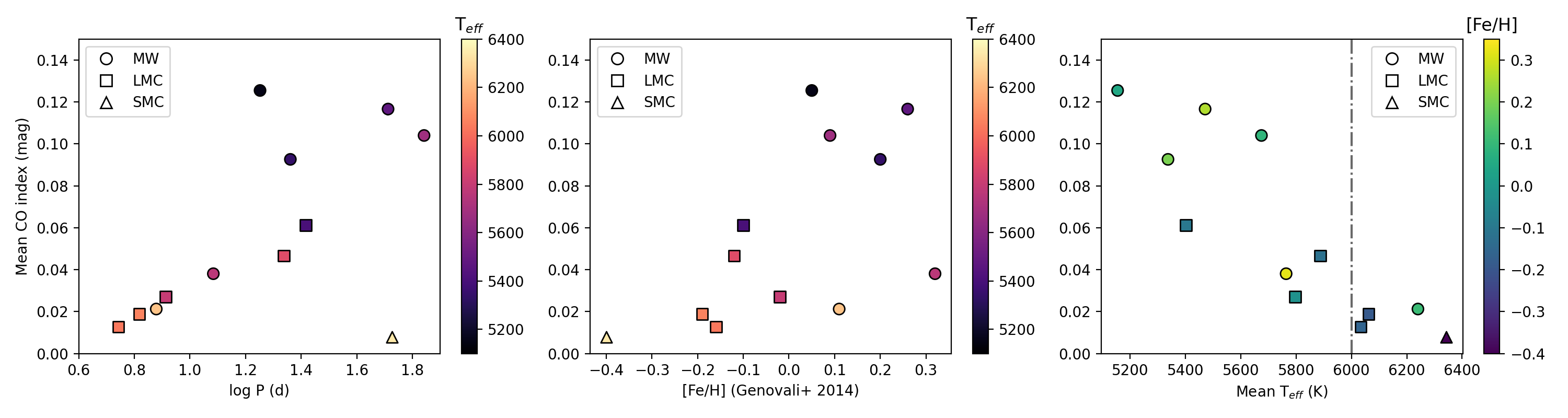}
    \caption{
    Mean value of the CO index as a function of $\log P$ (left), [Fe/H] (middle) and mean $T_{\text{eff}}$ (right). Symbols denote the "galaxy proxy" adopted for each Cepheid. The dot-dash line in the right panel indicates the CO destruction temperature of 6000~K. In the left and middle panels the points are colour-coded by mean $T_{\text{eff}}$, with points in the right hand panel coloured by \citet{genovali14} [Fe/H] values. The value for ET~Vul of [A/H]~$=-0.4$~dex comes from \citet{harris81}. The CO index appears to be stronger in Cepheids with higher metallicity, but the relationship has significant scatter. This scatter is believed to be due to the range of mean $T_{\text{eff}}$ across the sample, and the amplitude of the $T_{\text{eff}}$ variation for an individual Cepheid.
    }
    \label{fig:CO_Teff_period}
\end{figure*}

\begin{figure}
    \centering
    \includegraphics[width=\columnwidth]%{CO_logP_Teff_labelled.png}
    {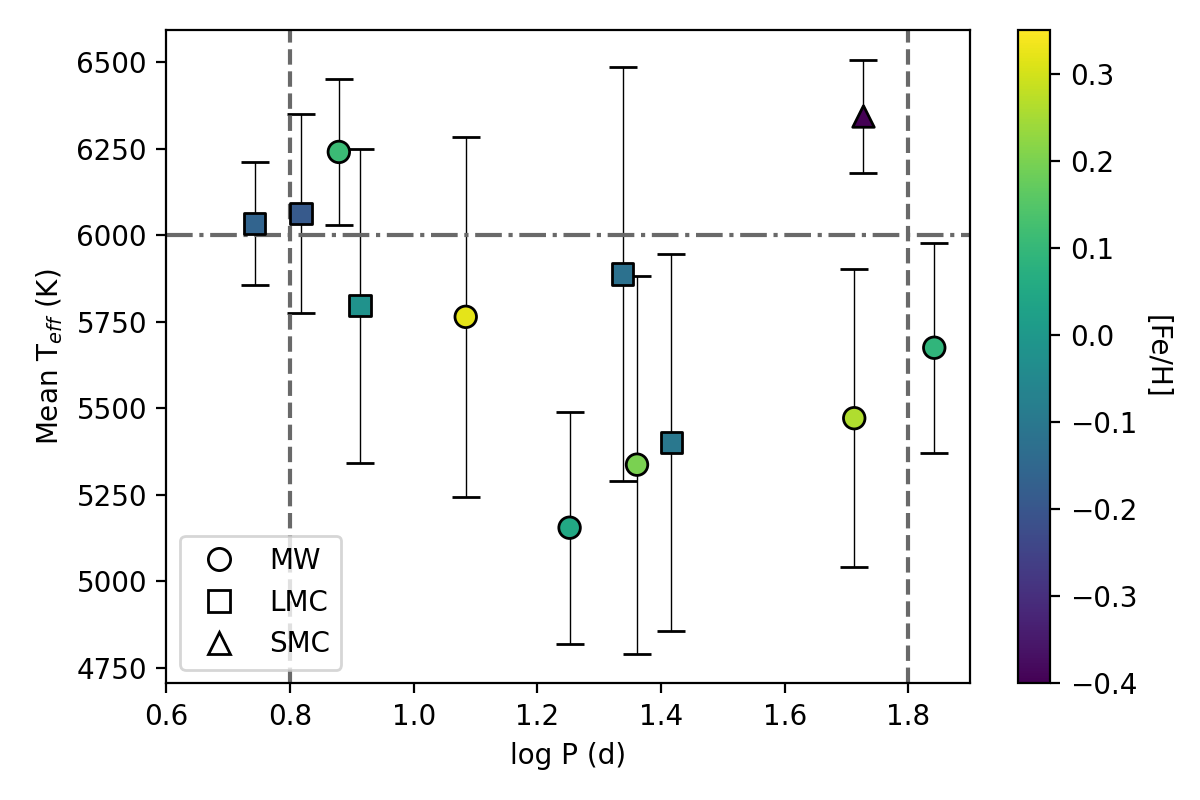}
    \caption{
    Mean $T_{\text{eff}}$ as a function of period. Error bars show the amplitude of the $T_{\text{eff}}$ variation found using \textsc{gloess} fitting. The dot-dash line at 6000~K indicates the expected CO destruction temperature. Dashed lines at $\log P =0.8$ and $\log P =1.8$ indicate the expected linear region of the mid-IR period-colour relation described in \citet{scowcroft16}. Points are colour coded by \citet{genovali14} [Fe/H] values, with symbols indicating the "galaxy proxy" adopted for each Cepheid. The value for ET~Vul of [A/H]~$=-0.4$~dex comes from \citet{harris81}.}
    \label{fig:t_amp}
\end{figure}

Figure~\ref{fig:CO_Teff_period} shows the relationship between the CO index, $T_{\text{eff}}$, [Fe/H], and period for our sample. For the purposes of this section, and to aid comparison with \citetalias{scowcroft16}, we have given each star in our sample a "galaxy proxy". We adopted mean metallicities of $+0.22\pm0.30$ (MW), $-0.22\pm0.19$ (LMC) and $-0.40\pm0.27$ (SMC) from \citet{2023A&A...671A.157H}, assuming Gaussian metallicity distributions about the mean values. The "galaxy proxy" of each Cepheid in our sample was set by finding the highest probability host galaxy at the Cepheid's metallicity. In figures~\ref{fig:CO_Teff_period} and ~\ref{fig:t_amp}, the three samples are represented by circles (MW), squares (LMC) and triangles (SMC).

The left and right panels of Figure~\ref{fig:CO_Teff_period} show the expected correlation between CO and $\log P$ (left) and mean $T_{\text{eff}}$ (right), with longer period, cooler Cepheids having stronger CO lines (with the exception of ET~Vul, see Section~\ref{sec:et_vul}). However, there is still noticeable dispersion, particularly in the cool, MW-metallicity Cepheids. The middle panel of Figure~\ref{fig:CO_Teff_period} shows that the mean value of the CO index has some dependency on [Fe/H], albeit with significant scatter.  We infer that the strength of the CO feature is driven by a combination of metallicity, mean $T_{\text{eff}}$, \emph{and} the amplitude of the $T_{\text{eff}}$ variation throughout the pulsation cycle.

Figure~\ref{fig:t_amp} shows the mean $T_{\text{eff}}$ value from the LTE models described in Section~\ref{sec:LTE}. Error bars indicate the amplitude of the $T_{\text{eff}}$ variation throughout the pulsation cycle, estimated from \textsc{gloess} fits. The $T_{\text{eff}}$ fits for each Cepheid are shown in Appendix~\ref{app:phase}, Figures~\ref{fig:svul} to \ref{fig:rzgem}. Figure~\ref{fig:t_amp} shows that the Cepheids in our sample with $\log~P \lesssim 0.9$ spend most of their time at temperatures above 6000~K where CO cannot form. The three Cepheids in this region have the very weak CO features, with little difference between the one LMC-like and two MW-like objects. This is consistent with the ``short period plateau'' region of the period-colour relation discussed in \citetalias{scowcroft16}.

As $\log~P$ increases, the \emph{mean} temperature of the Cepheid decreases and the CO strength increases. This is consistent with the main period-colour relation seen in \citetalias{scowcroft16}; the increase in CO and cooler average temperatures will induce a larger mid-IR colour variation. However, for Cepheids such as IY~Cyg ($\log P \approx 1.33$), which has a large $T_{\text{eff}}$ amplitude and LMC-like metallicity, a large fraction of the pulsation cycle is spent at $T_{\text{eff}} > 6000$~K, leading to weaker CO features. The effect is made clear when comparing with CP~Cep ($\log P \approx 1.25$); CP~Cep's shorter period would imply a higher $T_{\text{eff}}$ and therefore weaker CO. However, its smaller $T_{\text{eff}}$ amplitude and MW-like metallicity result in a higher mean CO strength, with less CO destruction at the hottest phases. This is further evidenced in Figure~\ref{fig:COandCN}, where we see that $\mathrm{CO}_{\text{min}}\approx 0.04$ for CP~Cep, rather than approaching zero as in most cases.

Figures~\ref{fig:CO_Teff_period} and \ref{fig:t_amp} show that both $T_{\text{eff}}$ and metallicity drive the strength of the CO feature, hence both parameters will impact the mid-IR colour. This means that in order to derive the metallicity of an individual Cepheid from its mid-IR colour, information about the star's temperature variation would also be required. In this work we have used KORG synthesised models along side our spectroscopic observations to derive $T_{\text{eff}}$ at each phase point (Section \ref{sec:LTE}). This is a reasonable approach for MW Cepheids; however, spectroscopic measurements quickly become prohibitively expensive for extragalactic Cepheids, particularly if multiple phase points are required. Alternative techniques to find the $T_{\text{eff}}$ variation, such as the the optical light curve decomposition technique from \citet{2010ApJ...719..335F} may solve this problem, allowing us to disentangle the temperature and metallicity effects sufficiently to provide a robust photometric metallicity indicator based on the CO feature. The Cepheid sample presented in this work does not cover a large enough range of metallicities for such an investigation, nor do all of the targets presented here have mid-IR data available. However, programs such as SAGE \citep{sage06}, and SH0ES \citep{riess22} contain extensive catalogues of mid-IR and optical photometry of extragalactic Cepheids. These large datasets of randomly sampled light curves can be supplemented with the templates from \citet{2025A&A...698A..97B} and \citet{chown21} and used in the optical light curve decomposition technique to disentangle the effects of temperature and metallicity, to calibrate a robust photometric metallicity indicator.

\section{Conclusions}
\label{sec:conclusions}

We have obtained medium resolution near-IR spectra for 12 Galactic Cepheids. Each star was observed at five or more different points during its pulsation cycle and six were observed at 10 or more. Our sample includes stars with periods from 5.5 to 69.5 days. We measured two carbon features, CN at 1.1-$\mu \rm m$ and CO at 2.3-$\mu \rm m$, and two hydrogen lines, Pa$\beta$ and Pa$\delta$. We performed LTE atmospheric fitting to metal lines in the $J$--band to obtain estimates of effective temperature, surface gravity and microturbulent velocity for each observation. The main results of this study are as follows:
\begin{enumerate}
    \item From the difference between CN and CO strengths, we identify a way to estimate abundance ratios of C, N, and O (e.g. [C/N]).
    \item The low-energy Paschen lines are affected by shocks propagating through the atmosphere and are not well represented by LTE estimates.
    \item The impact of the first dredge-up of processed material is much higher for CN than CO, consistent with previous studies of atomic CNO abundances.
\end{enumerate}

The potential effects of shocks on our measurements are explored, particularly on the lower energy Paschen series. We discuss the possibility that ET~Vul may be the result of a non-canonical formation channel, a binary merger, due to its nonconformity with expected parameters. We connect our near-IR measurements to the period-colour-metallicity relation in the mid-IR regime. Through this connection, we show that the amplitude of temperature variation, in addition to mid-IR colours, would be necessary to derive individual metallicities for extragalactic Cepheids.

The near-IR features of the molecule CO and radical CN can be used to explore temperature and composition of Cepheids. Their absorption bands are easily observable at medium (presented here) and low resolutions, making them attractive features for study in Cepheids with the instrument NIRSpec on \emph{JWST}.

\section*{Acknowledgements}
 
This work has made use of the VALD database, operated at Uppsala University, the Institute of Astronomy RAS in Moscow, and the University of Vienna. This paper includes data collected by the TESS mission, which are publicly available from the Mikulski Archive for Space Telescopes (MAST). Based on observations obtained with the Apache Point Observatory 3.5-meter telescope, which is owned and operated by the Astrophysical Research Consortium. We thank the anonymous referee for their comments and suggestions, which have improved this work.

This work was partially supported by David Nataf’s startup grant, specifically the University of Iowa’s Year 2 P3 Strategic Initiatives Program through funding received for the project entitled “High Impact Hiring Initiative (HIHI): A Program to Strategically Recruit and Retain Talented Faculty.

Software used in this work: \textsc{astropy} \citep{astropy13,astropy18,astropy22}, \textsc{numpy} \citep{numpy}, \textsc{scipy} \citep{scipy}

%%%%%%%%%%%%%%%%%%%%%%%%%%%%%%%%%%%%%%%%%%%%%%%%%%
\section*{Data Availability}

The data discussed in this paper will be made available upon reasonable request.

%%%%%%%%%%%%%%%%%%%% REFERENCES %%%%%%%%%%%%%%%%%%

% The best way to enter references is to use BibTeX:

\bibliographystyle{mnras}
\bibliography{_bibliography} % if your bibtex file is called _bibliography.bib

% Alternatively you could enter them by hand, like this:
% This method is tedious and prone to error if you have lots of references
%\begin{thebibliography}{99}
%\bibitem[\protect\citeauthoryear{Author}{2012}]{Author2012}
%Author A.~N., 2013, Journal of Improbable Astronomy, 1, 1
%\bibitem[\protect\citeauthoryear{Others}{2013}]{Others2013}
%Others S., 2012, Journal of Interesting Stuff, 17, 198
%\end{thebibliography}

%%%%%%%%%%%%%%%%%%%%%%%%%%%%%%%%%%%%%%%%%%%%%%%%%%

%%%%%%%%%%%%%%%%% APPENDICES %%%%%%%%%%%%%%%%%%%%%

%%%%%%%%%%%%%%%%%%%%%%%%%%%%%%%%%%%%%%%%%%%%%%%%%%
\appendix

\section{Asymmetry of Hydrogen Lines}
\label{app:asymmetry}
\begin{figure}
    \centering
    \includegraphics[width=\columnwidth]{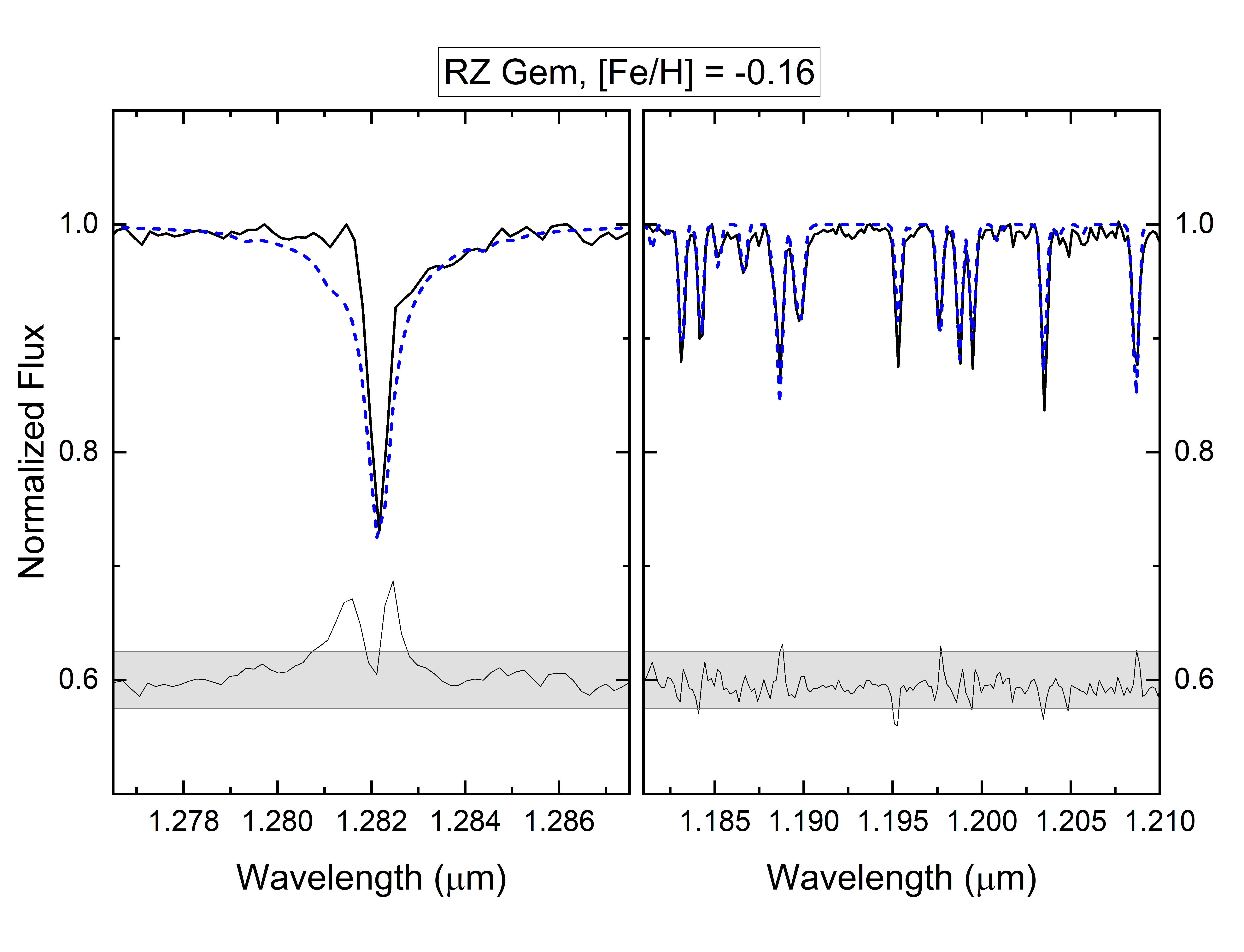}
    \caption{An example of asymmetry seen in the Pa$\beta$ line. Data for RZ~Gem (black solid line) and the best fit atmosphere model (blue dashed line). Left side shows the Pa$\mathrm{\beta}$ line and the right side a section of fitting region from Figure \ref{fig:fitting}. Residuals (observation $-$ model) are shown centred at normalized flux of 0.6, with the shaded area representing $\pm 0.025$. The synthetic model parameters are $T_{\rm eff} = 6039$ K, log $g = 1.87$, and $v_{\rm mic} = 4.7$ km s$^{-1}$.}
    \label{fig:asymm}
\end{figure}

\begin{figure}
    \centering
    \includegraphics[width=\columnwidth]{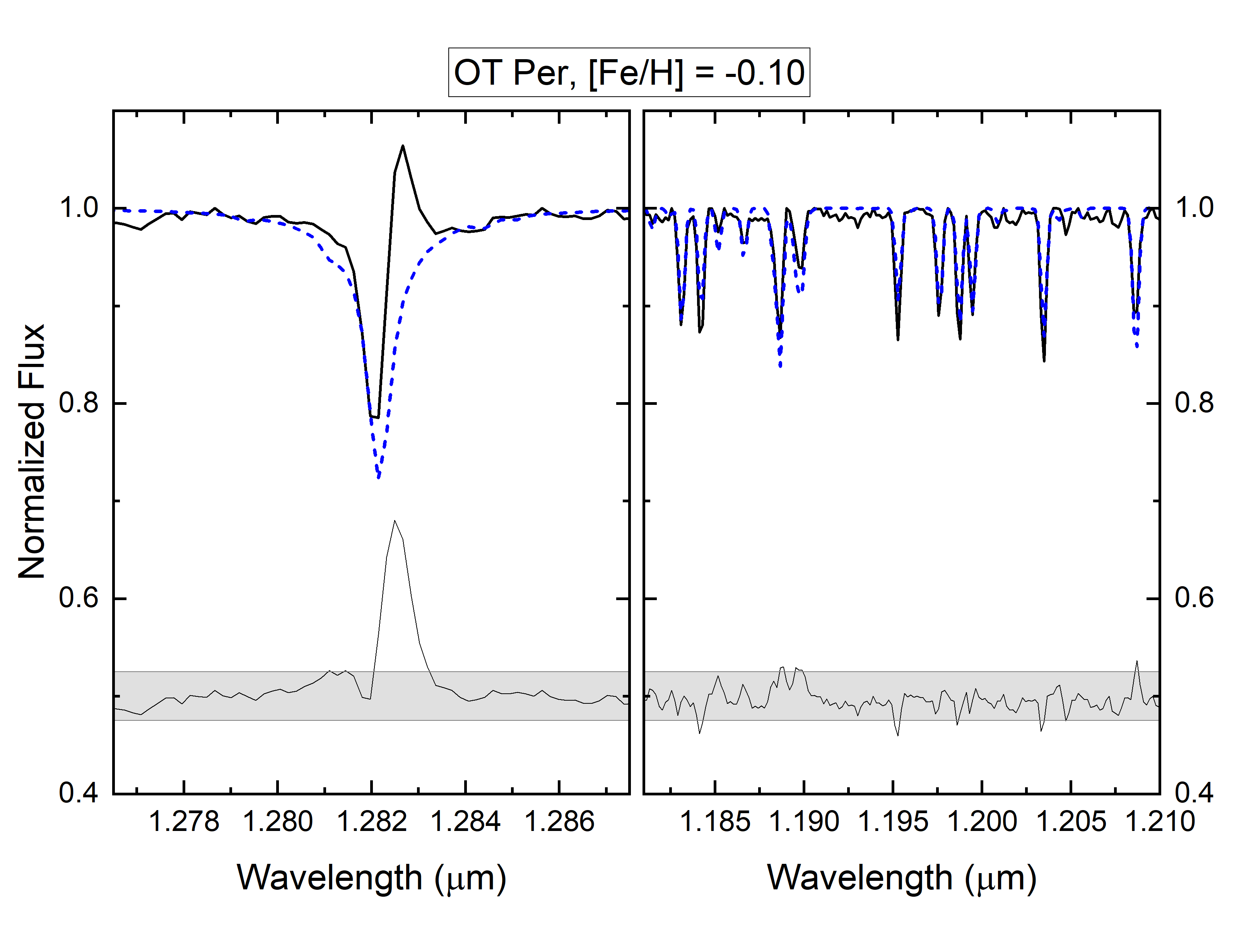}
    \caption{Another example of asymmetry seen in the Pa$\beta$ line. Data for OT~Per (black solid line) and the best fit atmosphere model (blue dashed line). Left side shows the Pa$\mathrm{\beta}$ line and the right side a section of fitting region from Figure \ref{fig:fitting}. Residuals (observation $-$ model) are shown centred at normalized flux of 0.5, with the shaded area representing $\pm 0.025$. The synthetic model parameters are $T_{\rm eff} = 5969$ K, log $g = 1.40$, and $v_{\rm mic} = 4.9$ km s$^{-1}$.}
    \label{fig:otper_shock}
\end{figure}

\section{Phase Results}
\label{app:phase}

\begin{table*}
    \centering
    \begin{tabular}{c|c|c|c|c|c|c|c|c|c|c|c|c|c|c}
        Star & Phase & HJD & $\chi_{\rm red}^{2}$ & $T_{\rm eff}$ & $\log g$ & $v_{\rm mic}$ & $\mathrm{CO}$ & $\sigma_{\rm CO}$ & $\mathrm{CN}$ & $\sigma_{\rm CN}$ & Pa$\mathrm{\beta}$ & $\sigma_{\rm Pa_\beta}$ & Pa$\mathrm{\delta}$ & $\sigma_{\rm Pa_\delta}$ \\
        &  & +2400000 & & (K) & ([cm s$^{-2}$]) & (km s$^{-1}$)& (mag)  & (mag) & (mag) & (mag) & (mag) & (mag) & (mag) & (mag) \\
        \hline 
        BM PER & 0.876 & 60324.590 & 35  & 5305.5 & 0.88 & 3.9 & 0.060 & 0.020 & 0.068 & 0.018 & 0.052 & 0.017 & 0.028 & 0.018 \\
        BM PER & 0.835 & 60346.605 & 128 & 5483.1 & 0.83 & 5.0 & 0.091 & 0.016 & 0.072 & 0.020 & 0.022 & 0.017 & 0.015 & 0.020 \\ 
        BM PER & 0.445 & 60360.614 & 129 & 4969.3 & 0.40 & 2.9 & 0.144 & 0.016 & 0.097 & 0.021 & 0.070 & 0.027 & 0.008 & 0.021 \\
        BM PER & 0.012 & 60373.639 & 345 & 6224.0 & 1.91 & 3.2 & 0.000 & 0.031 & 0.006 & 0.022 & 0.156 & 0.011 & 0.110 & 0.022 \\
        BM PER & 0.317 & 60380.647 & 228 & 5290.3 & 0.94 & 3.5 & 0.084 & 0.017 & 0.063 & 0.033 & 0.045 & 0.017 & 0.029 & 0.033 \\
        BM PER & 0.820 & 60530.001 & 69 & 5361.7 & 0.76 & 4.7 & 0.107 & 0.031 & 0.077 & 0.034 & 0.021 & 0.022 & 0.013 & 0.034 \\
        BM PER & 0.824 & 60576.011 & 96 & 5319.6 & 0.63 & 4.5 & 0.082 & 0.015 & 0.074 & 0.023 & 0.002 & 0.016 & 0.008 & 0.023 \\
        BM PER & 0.997 & 60580.000 & 404 & 4957.8 & 0.74 & 1.7 & -0.002 & 0.018 & -0.005 & 0.021 & 0.142 & 0.020 & 0.107 & 0.021\\
        BM PER & 0.781 & 60598.010 & 46 & 5181.1 & 0.65 & 4.2 & 0.177 & 0.021 & 0.125 & 0.021 & 0.028 & 0.023 & 0.013 & 0.021 \\
        BM PER & 0.130 & 60606.016 & 142 & 6200.7 & 1.52 & 5.3 & -0.001 & 0.018 & -0.002 & 0.019 & 0.081 & 0.012 & 0.056 & 0.019 \\
        ... & ... & ... & ... & ... & ... & ... & ... & ... & ... & ... & ... & ... & ... & ...
    \end{tabular}
    \caption{Per-phase data for each Cepheid in the sample. The reported phases and HJD are the mean for the exposure sequences. The best-fit parameters ($T_{\rm eff}$, $\log g$, $v_{\rm mic}$) from the KORG fitting routine and the reduced-$\chi^2$ are given. The band and line index magnitudes for CO, CN, Pa$\beta$, and Pa$\delta$ and their RMS uncertainties are also shown. This represents a portion of the full data set, and the table in its entirety is available via the online version of the journal.}
    \label{tab:placeholder}
\end{table*}

\begin{figure}
    \centering
    \includegraphics[width=\columnwidth]{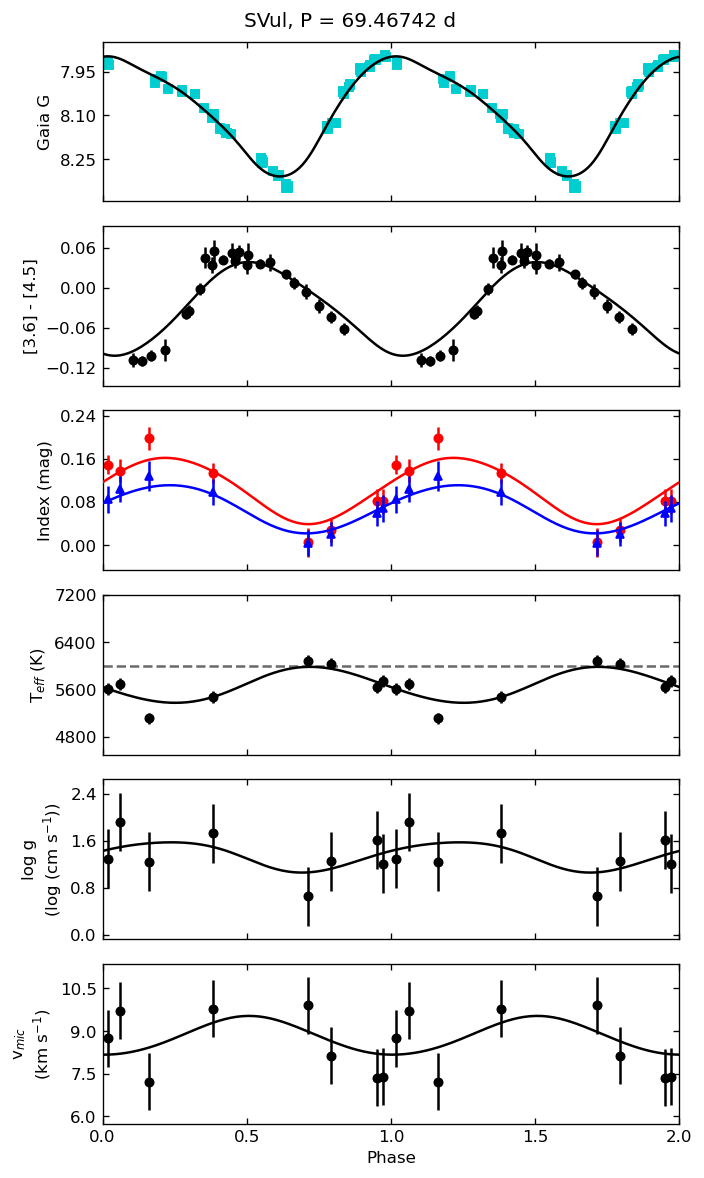}
    \caption{From top to bottom: $G$ light curve from \emph{Gaia}, mid-IR [3.6]-[4.5] colour curve \citep{monson12}, CO and CN values as in Figure \ref{fig:COandCN}, $T_{\rm eff}$, $\mathrm{log\,}g$, and $v_{\rm mic}$ from KORG fitting routine. Solid lines denote \textsc{gloess} curves. A dashed line is shown in the $T_{\rm eff}$ panel at $6000$~K to indicate the approximate destruction temperature of CO.  }
    \label{fig:svul}
\end{figure}

\begin{figure}
    \centering
    \includegraphics[width=\columnwidth]{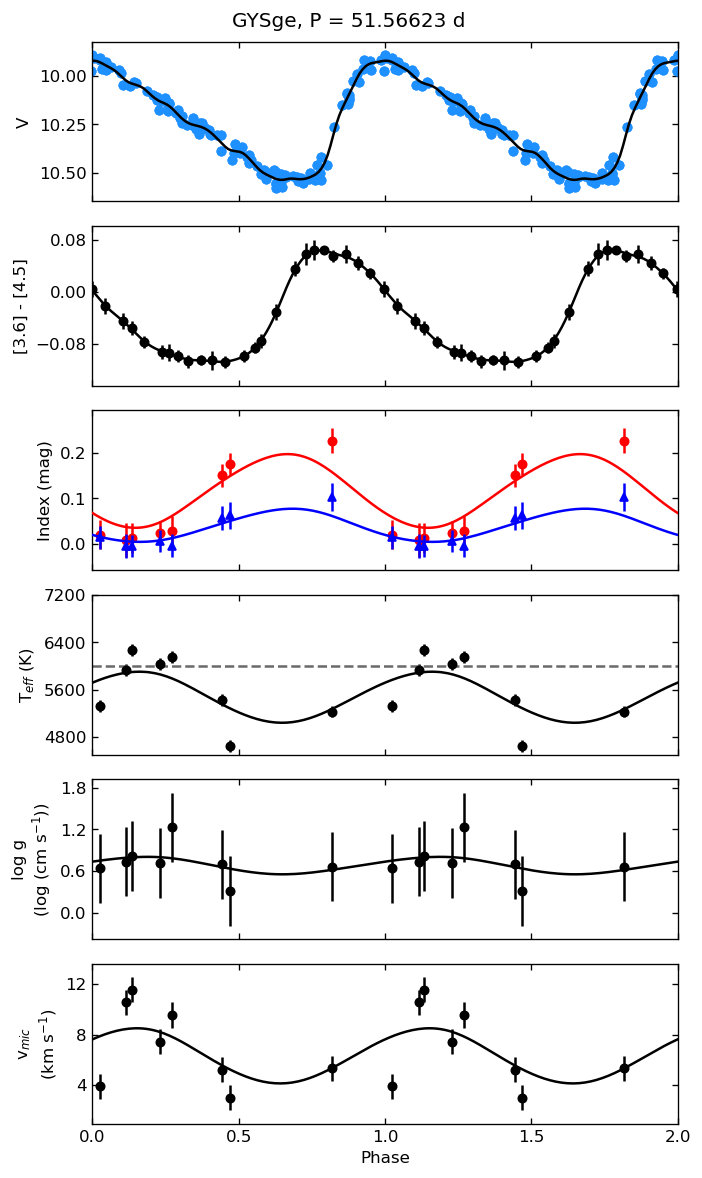}
    \caption{The same as Figure \ref{fig:svul} for GY~Sge, but the $V$ light curve is from ASAS-SN \citep{shappee14,kochanek17}.}
    \label{fig:gysge}
\end{figure}

\begin{figure}
    \centering
    \includegraphics[width=\columnwidth]{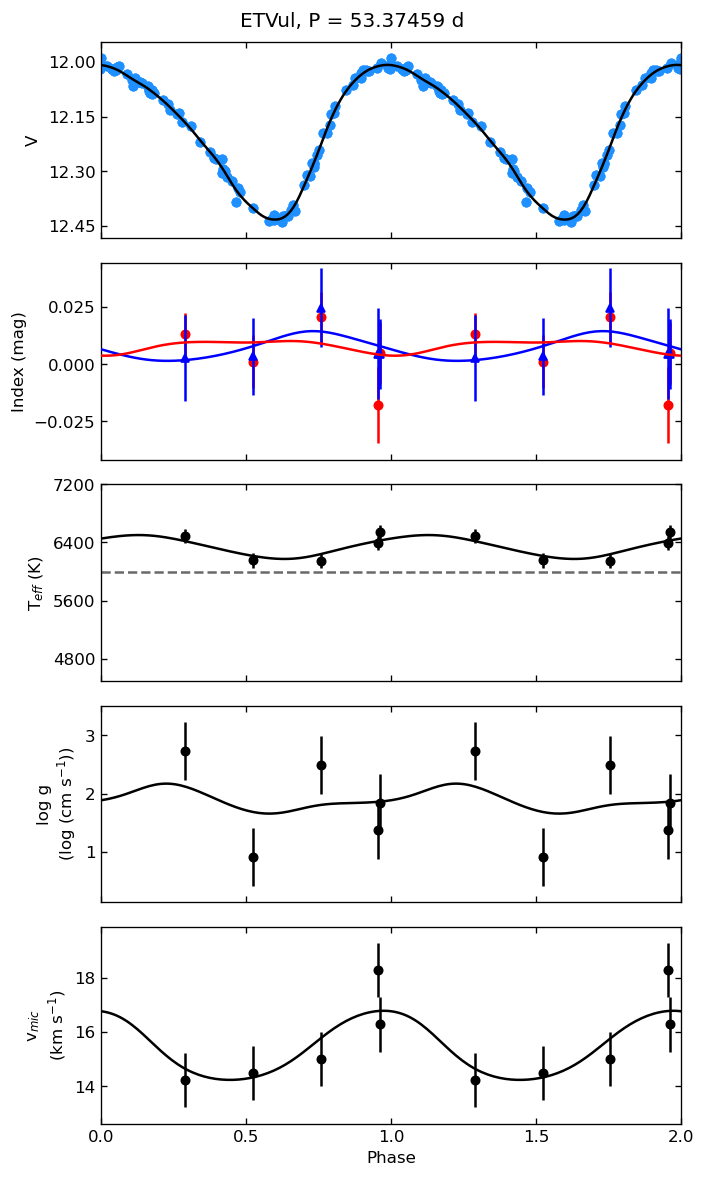}
    \caption{From top to bottom: $V$ light curve from ASAS-SN for ET~Vul, CO and CN values as in Figure \ref{fig:COandCN}, phase $T_{\rm eff}$ results, phase $\mathrm{log\,}g$ results, and $v_{\rm mic}$ from KORG fitting routine. Solid lines denote \textsc{gloess} curves. A dashed line is shown in the $T_{\rm eff}$ panel at $6000$~K to indicate the approximate destruction temperature of CO.}
    \label{fig:etvul}
\end{figure}

\begin{figure}
    \centering
    \includegraphics[width=\columnwidth]{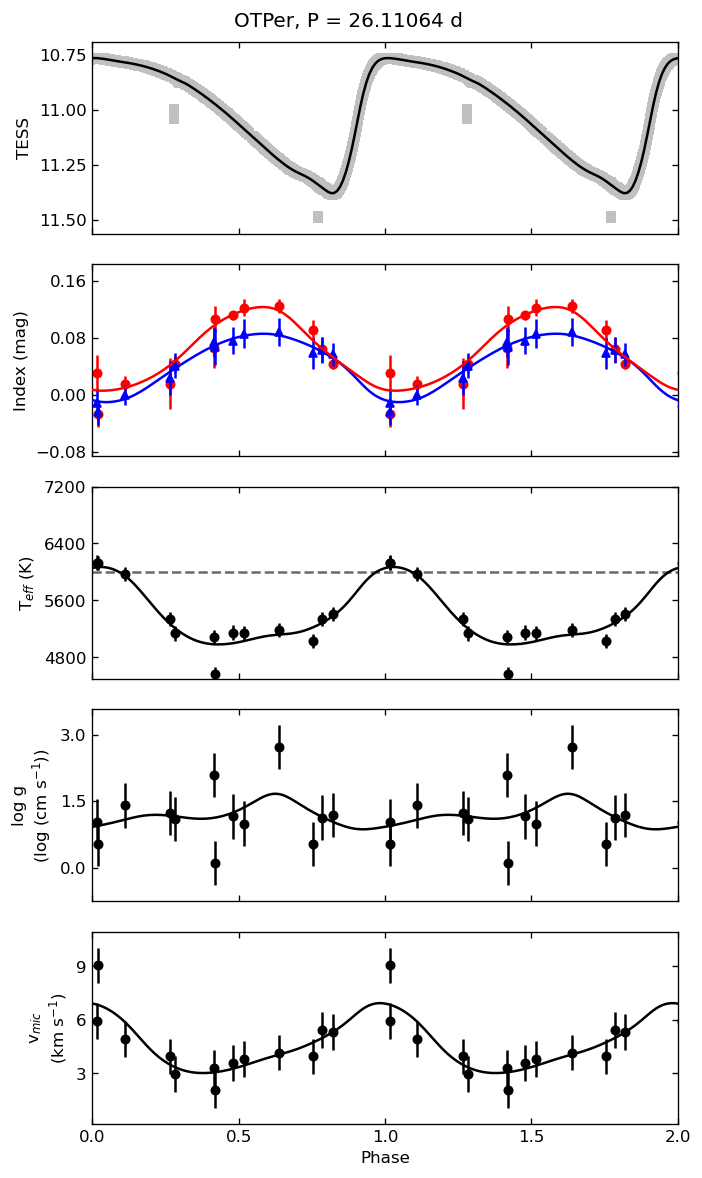}
    \caption{The same as Figure \ref{fig:otper} for OT~Per, but the TESS light curve is in the top panel.}
    \label{fig:otper}
\end{figure}

\begin{figure}
    \centering
    \includegraphics[width=\columnwidth]{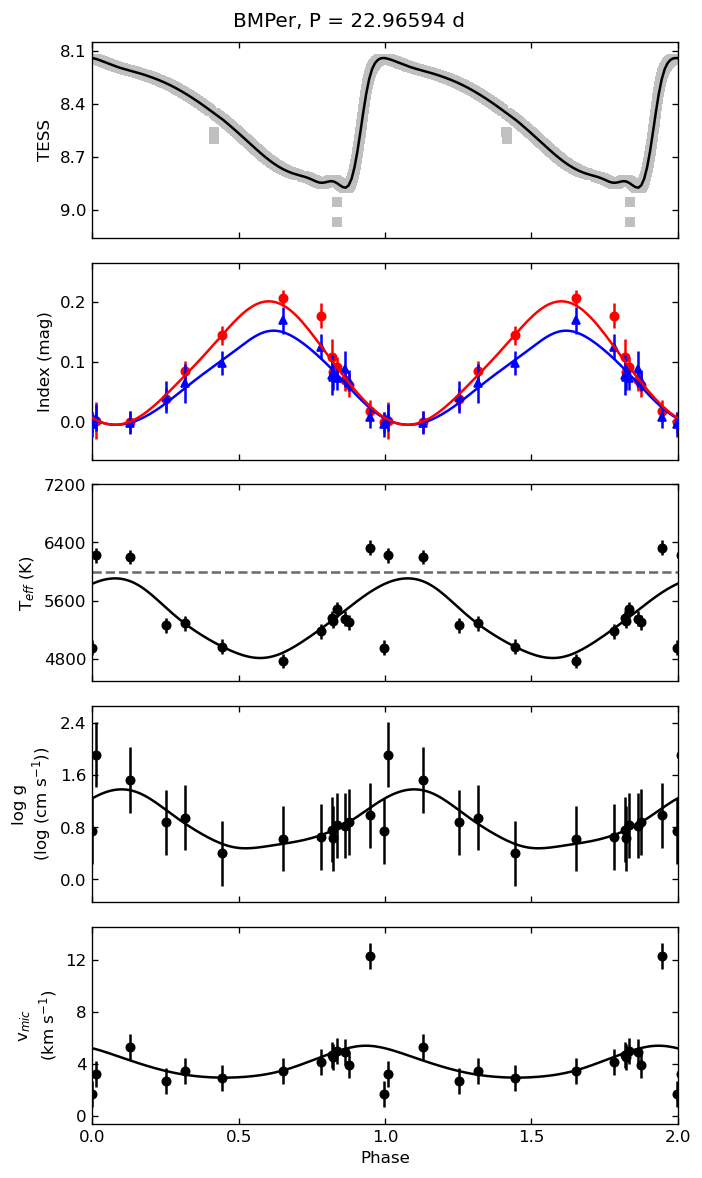}
    \caption{The same as Figure \ref{fig:otper} for BM~Per.}
    \label{fig:bmper}
\end{figure}

\begin{figure}
    \centering
    \includegraphics[width=\columnwidth]{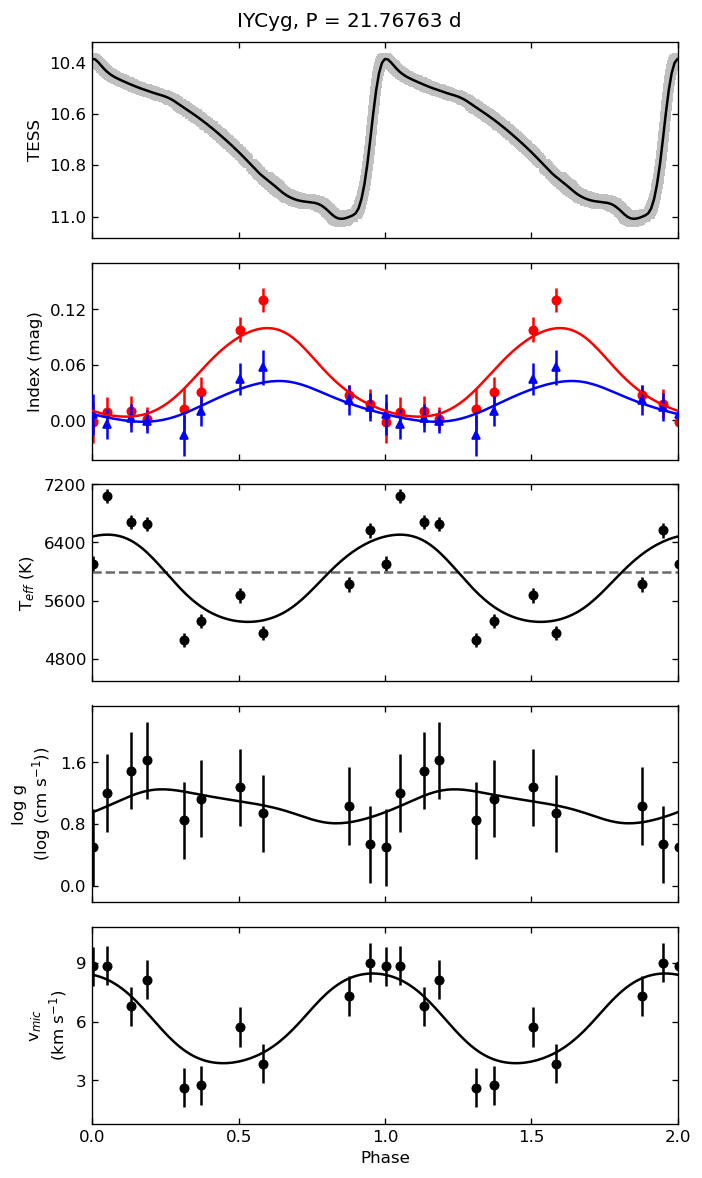}
     \caption{The same as Figure \ref{fig:otper} for IY~Cyg.}
    \label{fig:iycyg}
\end{figure}

\begin{figure}
    \centering
    \includegraphics[width=\columnwidth]{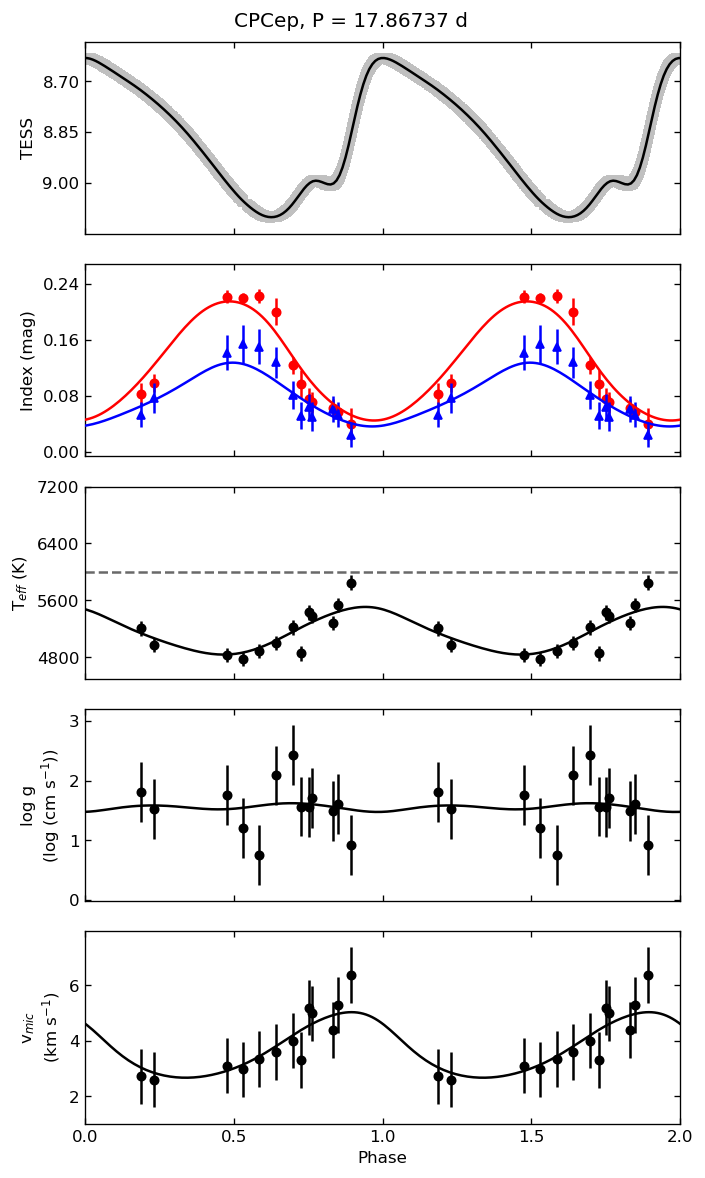}
    \caption{The same as Figure \ref{fig:otper} for CP~Cep.}
    \label{fig:cpcep}
\end{figure}

\begin{figure}
    \centering
    \includegraphics[width=\columnwidth]{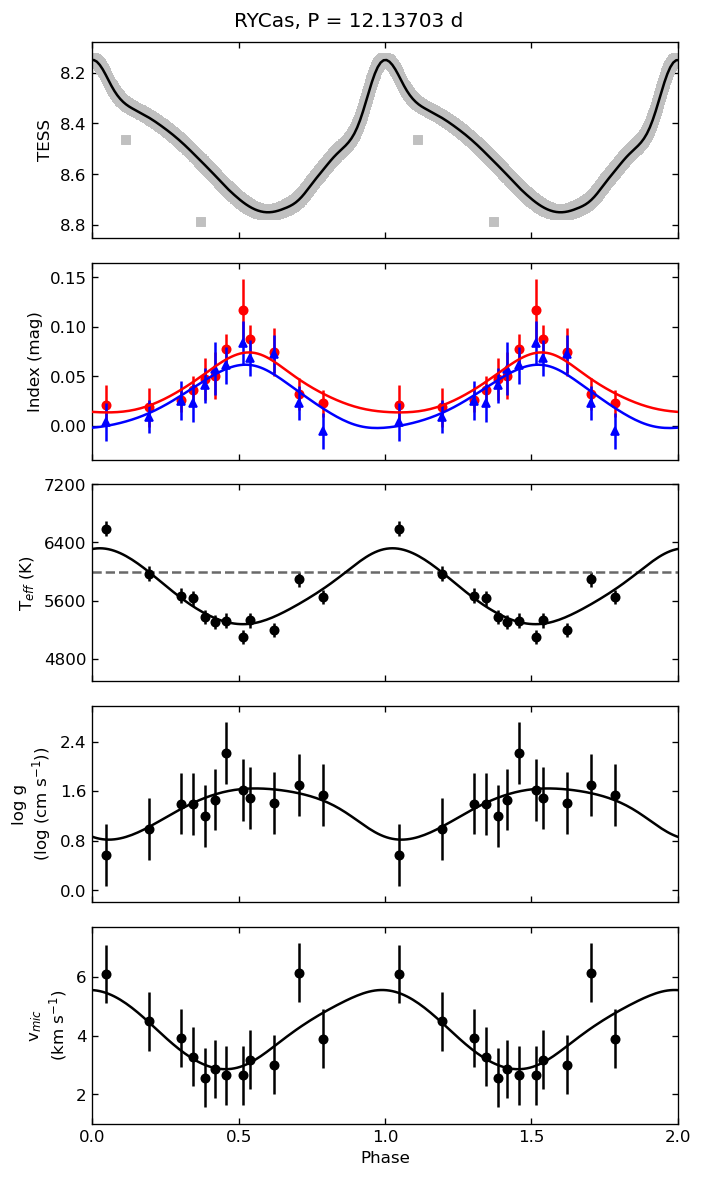}
    \caption{The same as Figure \ref{fig:otper} for RY~Cas.}
    \label{fig:rycas}
\end{figure}

\begin{figure}
    \centering
    \includegraphics[width=\columnwidth]{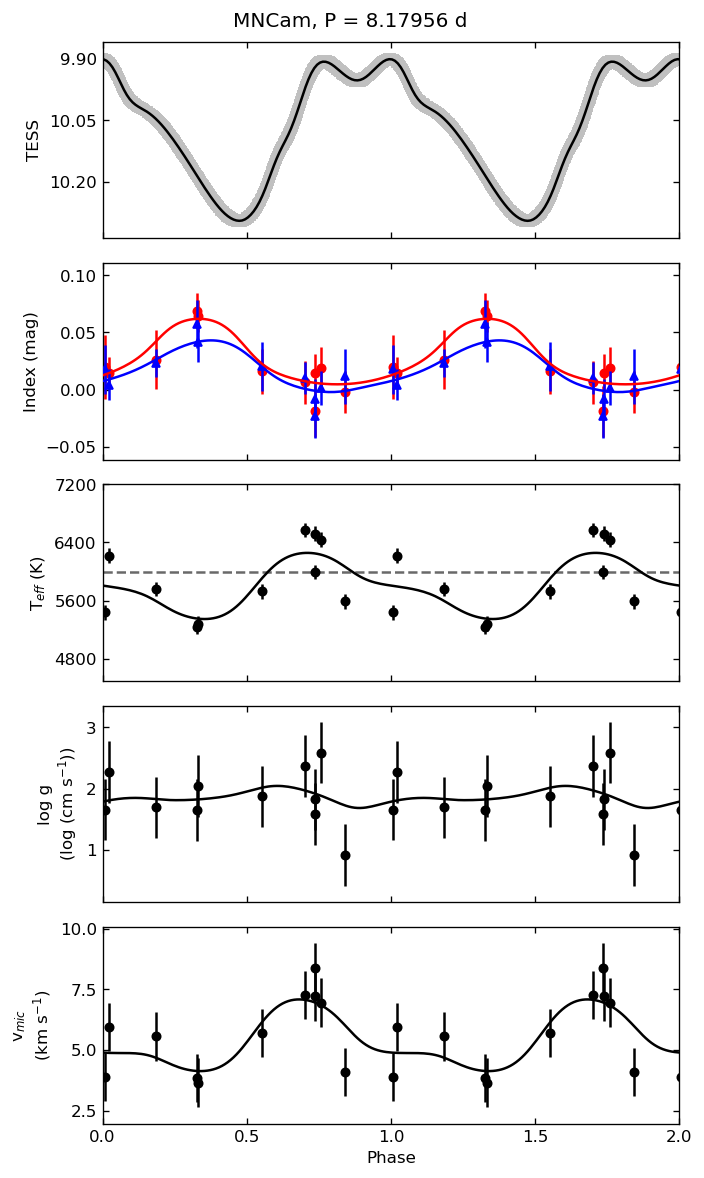}
    \caption{The same as Figure \ref{fig:otper} for MN~Cam.}
    \label{fig:mncam}
\end{figure}

\begin{figure}
    \centering
    \includegraphics[width=\columnwidth]{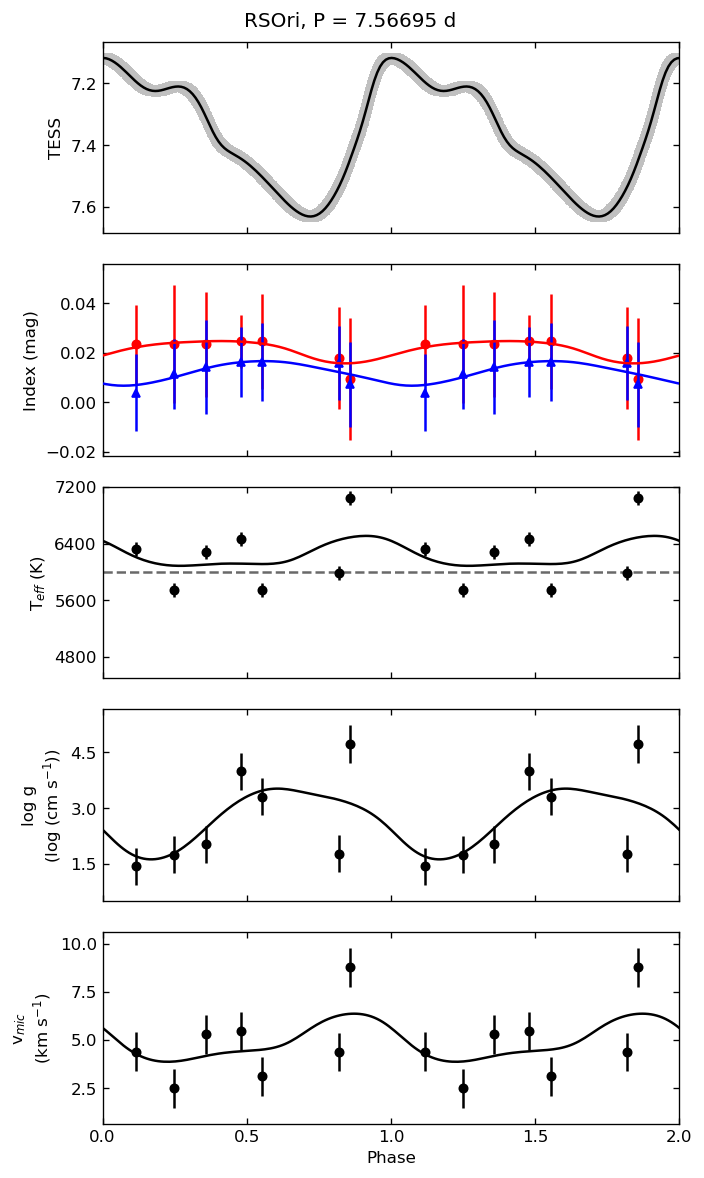}
    \caption{The same as Figure \ref{fig:otper} for RS~Ori.}
    \label{fig:rsori}
\end{figure}

\begin{figure}
    \centering
    \includegraphics[width=\columnwidth]{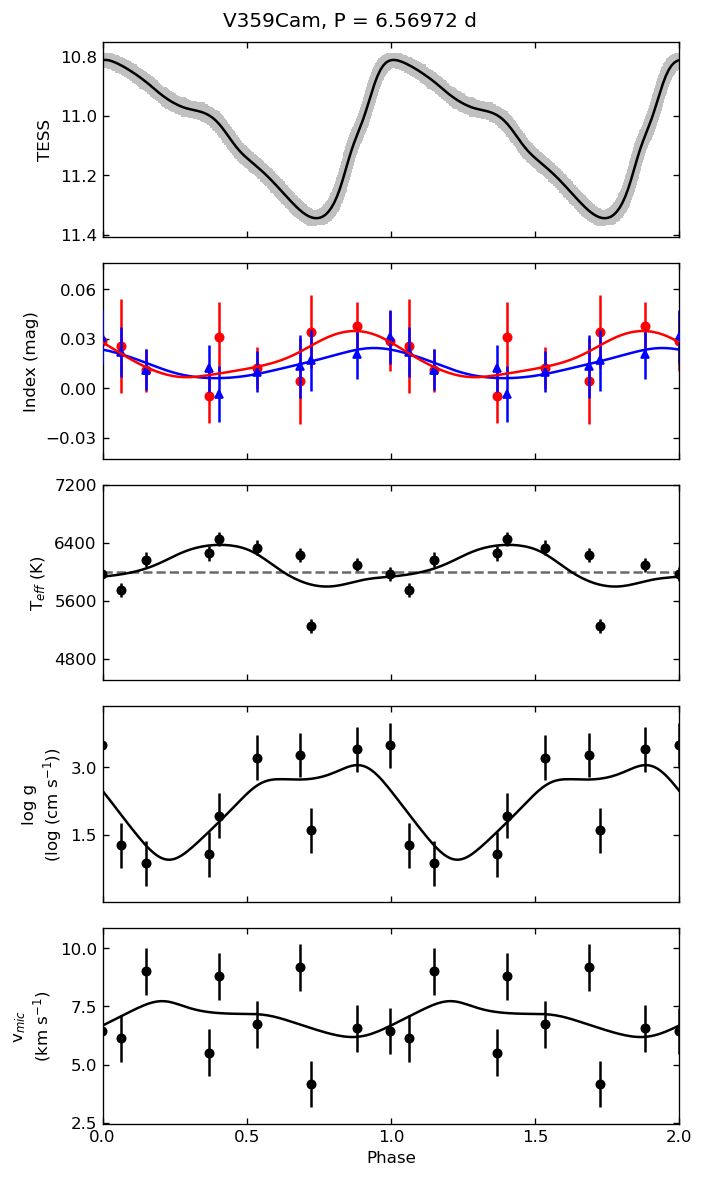}
    \caption{The same as Figure \ref{fig:otper} for V359~Cam.}
    \label{fig:v359cam}
\end{figure}

\begin{figure}
    \centering
    \includegraphics[width=\columnwidth]{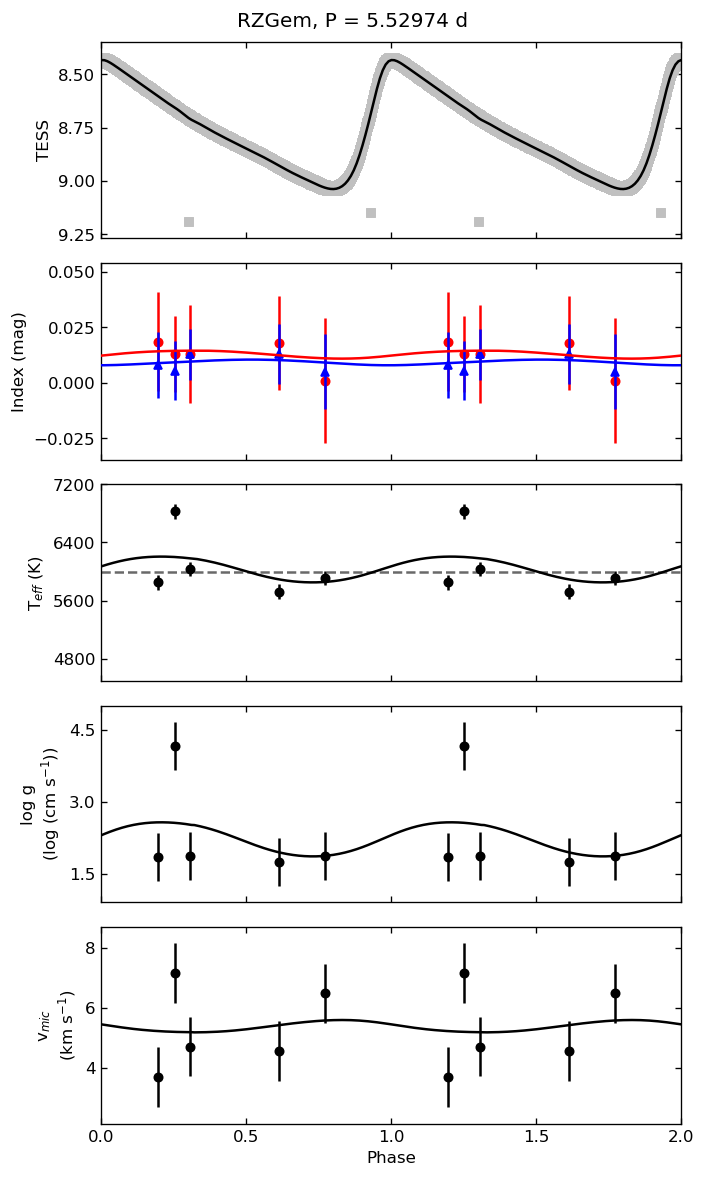}
    \caption{The same as Figure \ref{fig:otper} for RZ~Gem.}
    \label{fig:rzgem}
\end{figure}

% Don't change these lines
\bsp	% typesetting comment
\label{lastpage}
\end{document}